\documentclass[10pt,conference]{IEEEtran}

\usepackage{fancyhdr}					
\usepackage{glossaries}                 
\usepackage{amsfonts}                   
\usepackage{pifont}                     
\usepackage{color-edits}                
\usepackage{xcolor}                     
\usepackage{enumitem}                   
\usepackage{subcaption}                 
\usepackage{graphicx}                   
\usepackage{makecell}                   
\usepackage{booktabs}                   
\usepackage{multirow}                   
\usepackage[]{algorithm2e}              
\usepackage{float}                      
\usepackage{dblfloatfix}                

\PassOptionsToPackage{hyphens}{url}\usepackage{hyperref}

\usepackage{cleveref}                   

\crefname{algocf}{Alg.}{Algs.}
\crefname{figure}{Fig.}{Figs.}
\crefname{equation}{Eq.}{Eqs.}

\newcommand{\hpcayear}{2025}

\setacronymstyle{long-short}
\newacronym{ici}{ICI}{inter-chiplet interconnect}
\newacronym{noc}{NoC}{network-on-chip}
\newacronym{srams}{SRAMs}{static random access memories}
\newacronym{c4}{C4}{controlled collapse chip connection}
\newacronym{pcb}{PCB}{printed circuit board}
\newacronym{d2d}{D2D}{die-to-die}
\newacronym{beol}{BEOL}{back end of line}
\newacronym{feol}{FEOL}{front end of line}
\newacronym{Tsvs}{TSVs}{Through-silicon vias}
\newacronym{phys}{PHYs}{physical layers}
\newacronym{emib}{EMIB}{embedded multi-die interconnect bridge}
\newacronym{dbhi}{DBHi}{direct bonded heterogeneous integration}
\newacronym{c2c}{C2C}{compute-to-compute}
\newacronym{c2m}{C2M}{compute-to-memory}
\newacronym{c2i}{C2I}{compute-to-IO}
\newacronym{m2i}{M2I}{memory-to-IO}
\newacronym{spm}{SPM}{scratchpad memory}
\newacronym{ucie}{UCIe}{universal chiplet interconnect express}
\newacronym{bow}{BoW}{bunch of wires}
\newacronym{mst}{MST}{minimum spanning tree}
\newacronym{br}{BR}{best random}
\newacronym{ga}{GA}{genetic algorithm}
\newacronym{sa}{SA}{simulated annealing}
\newacronym{twl}{TWL}{total wire length}
\newacronym{cowos}{CoWoS}{Chip-on-Wafer-on-Substrate}
\newacronym{vc}{VC}{virtual channel}
\newacronym{vlsi}{VLSI}{very large scale integration}

\newif\ifnb     
\nbtrue

\newif\ifcom    
\comfalse

\newif\ifps     
\psfalse

\newcommand{\name}{PlaceIT}

\addauthor[Summary]{ps}{orange}
\addauthor[Patrick]{pi}{teal}
\addauthor[Maciej]{mb}{blue}
\addauthor[Torsten]{th}{red}
\addauthor[L. Benini]{lb}{violet}

\newcommand{\ps}[1]{\ifps\pscomment{#1}\fi}
\renewcommand{\pi}[1]{\ifcom\picomment{#1}\fi}

\setlist{leftmargin=1em}

\newcommand{\cmark}{\ding{52}}%
\newcommand{\xmark}{\ding{56}}%
\newcommand{\qmark}{\raisebox{-0.25em}{\includegraphics[height=1em]{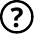}}}%

\newcommand{\hpcasubmissionnumber}{279}
\title{PlaceIT: \underline{Place}ment-based Inter-Chiplet\\\underline{I}nterconnect \underline{T}opologies}

\ifnb
\def\hpcacameraready{} 
\fi

\newcommand\hpcaauthors{Patrick Iff$^\dagger$ and Benigna Bruggmann$^\dagger$ and Maciej Besta$^\dagger$ and Luca Benini$^{\dagger\ddagger}$ and Torsten Hoefler$^\dagger$}
\newcommand\hpcaaffiliation{ETH Zurich, Zurich, Switzerland$^\dagger$, University of Bologna, Italy$^\ddagger$}
\newcommand\hpcaemail{patrick.iff@inf.ethz.ch, maciej.besta@inf.ethz.ch, torsten.hoefler@inf.ethz.ch, lbenini@iis.ee.ethz.ch}

\author{
  \ifdefined\hpcacameraready
    \IEEEauthorblockN{\hpcaauthors{}}
      \IEEEauthorblockA{
        \hpcaaffiliation{} \\
        \hpcaemail{}
      }
  \else
    \IEEEauthorblockN{\normalsize{HPCA \hpcayear{} Submission
      \textbf{\#\hpcasubmissionnumber{}}} \\
      \IEEEauthorblockA{
        Confidential Draft \\
        Do NOT Distribute!!
		\vspace{-0.75em}
      }
    }
  \fi 
}

\fancypagestyle{camerareadyfirstpage}{%
  \fancyhead{}
  
  \fancyhead[C]{
    \ifdefined\aeopen
    \parbox[][12mm][t]{13.5cm}{}    
    \else
      \ifdefined\aereviewed
      \parbox[][12mm][t]{13.5cm}{}
      \else
      \ifdefined\aereproduced
      \parbox[][12mm][t]{13.5cm}{}
      \else
      \parbox[][0mm][t]{13.5cm}{}
    \fi 
    \fi 
    \fi 
    \ifdefined\aeopen 
      \includegraphics[width=12mm,height=12mm]{ae-badges/open-research-objects.pdf}
    \fi 
    \ifdefined\aereviewed
      \includegraphics[width=12mm,height=12mm]{ae-badges/research-objects-reviewed.pdf}
    \fi 
    \ifdefined\aereproduced
      \includegraphics[width=12mm,height=12mm]{ae-badges/results-reproduced.pdf}
    \fi
  }
  \fancyfoot[C]{}
}
\begin{document}
\maketitle

\ifdefined\hpcacameraready 
  \thispagestyle{camerareadyfirstpage}
  \pagestyle{empty}
\else
  \thispagestyle{plain}
  \pagestyle{plain}
\fi

\newcommand{\hpcaheight}{0mm}
\ifdefined\eaopen
\renewcommand{\hpcaheight}{12mm}
\fi


\begin{abstract}
2.5D integration technology is gaining traction as it copes with the exponentially growing design cost of modern integrated circuits. 
A crucial part of a 2.5D stacked chip is a low-latency and high-throughput inter-chiplet interconnect (ICI).
Two major factors affecting the latency and throughput are the topology of links between chiplets and the chiplet placement.
In this work, we present \name, a novel methodology to jointly optimize the ICI topology and the chiplet placement.
While state-of-the-art methods optimize the chiplet placement for a predetermined ICI topology, or they select one topology out of a set of candidates, we generate a completely new topology for each placement.
Our process of inferring placement-based ICI topologies connects chiplets that are in close proximity to each other, making it particularly attractive for chips with silicon bridges or passive silicon interposers with severely limited link lengths.
We provide an open-source implementation of our method that optimizes the placement of homogeneously or heterogeneously shaped chiplets and the ICI topology connecting them for a user-defined mix of four different traffic types.
We evaluate our methodology using synthetic traffic and traces, and we compare our results to a 2D mesh baseline.
\name~reduces the latency of synthetic L1-to-L2 and L2-to-memory traffic, the two most important types for cache coherency traffic, by up to $28$\% and $62$\%, respectively.
It also achieve an average packet latency reduction of up to $18\%$ on traffic traces.
\name~ enables the construction of 2.5D stacked chips with low-latency ICIs.

\end{abstract}

\ifnb
\begin{center}
\vspace{-0.25em}
\textbf{Website \& code:} \url{https://github.com/spcl/placeit}
\vspace{-0.5em}
\pi{\\ToDo: Create repo, double-check URL}
\end{center}
\else
\begin{center}
\vspace{-0.25em}
\textbf{Code:} \url{https://www.dropbox.com/scl/fi/jzwxdm1s58ml1inw43uw4/PlaceIT.zip?rlkey=ue12hxwlskkyem12dyqtcijhc&st=6s0ewq7k&dl=0}
\vspace{-0.5em}
\pi{\\ToDo: Create repo, double-check URL}
\end{center}
\fi

\begin{figure*}[h]
\centering
\captionsetup{justification=centering}
\begin{subfigure}{0.22 \textwidth}
\centering
\includegraphics[width=1.0\textwidth]{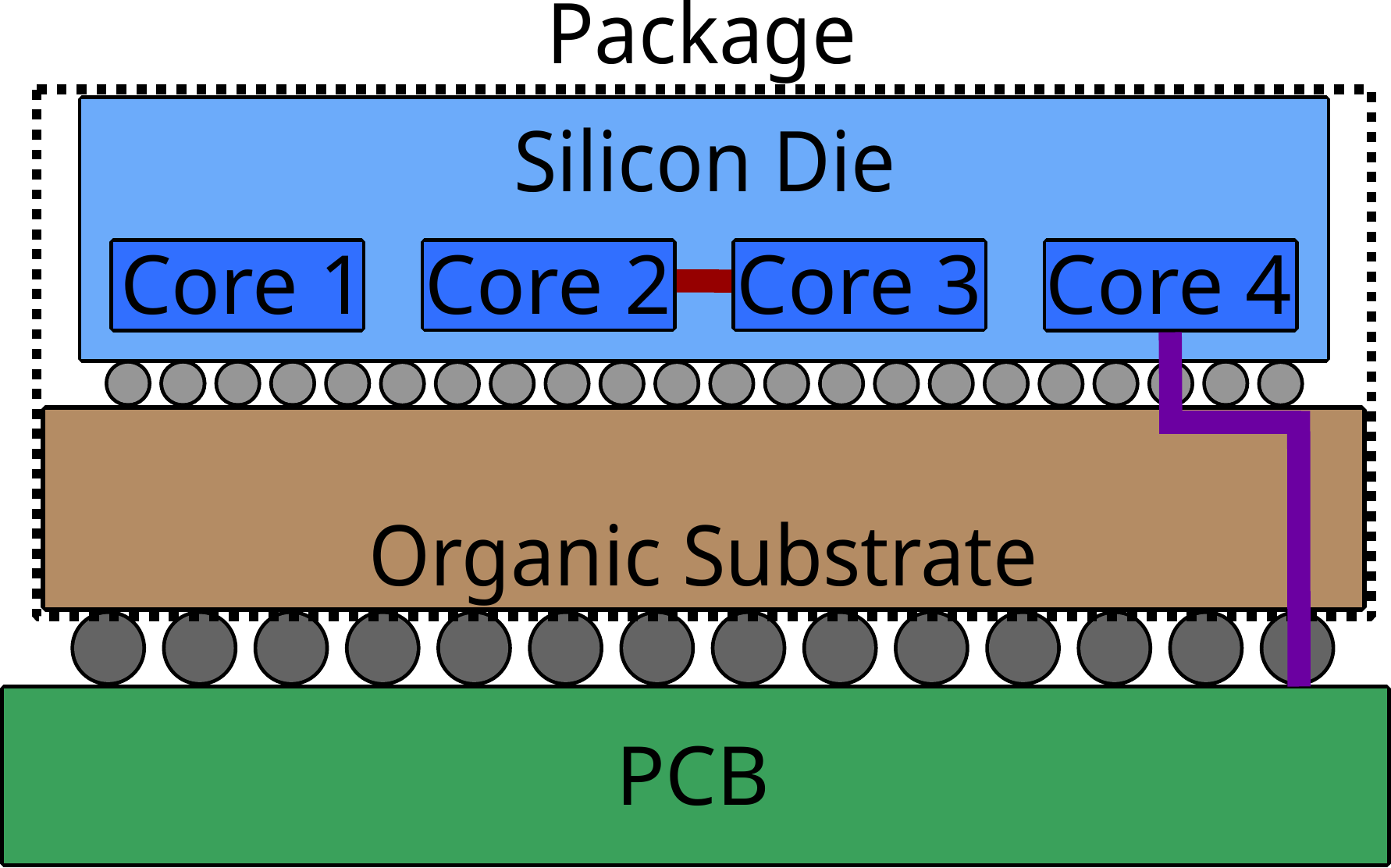}
\caption{Monolithic chip.}
\label{fig:back-monolithic}
\end{subfigure}
\begin{subfigure}{0.22 \textwidth}
\centering
\includegraphics[width=1.0\textwidth]{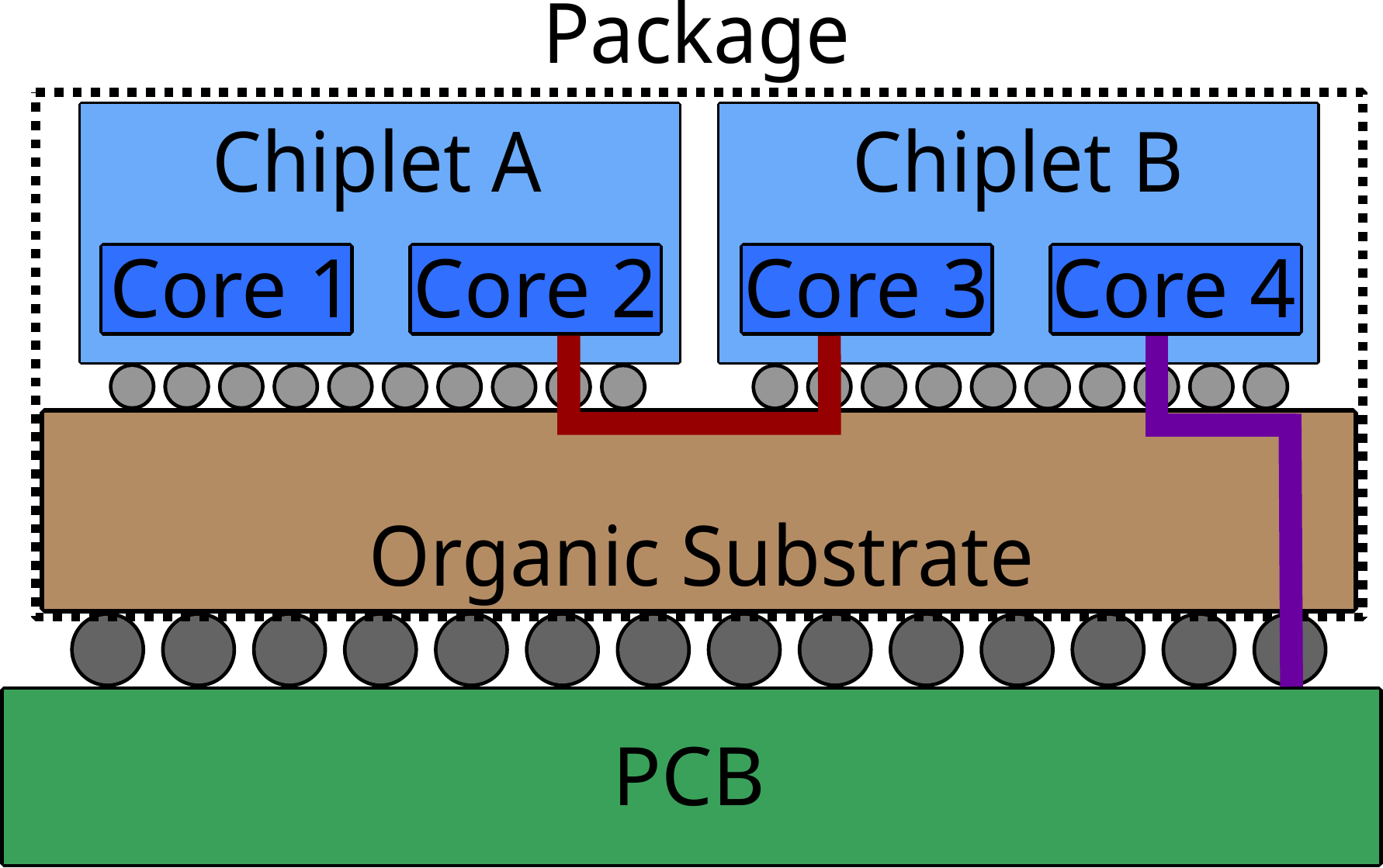}
\caption{Organic substrate.}
\label{fig:back-substrate}
\end{subfigure}
\begin{subfigure}{0.22 \textwidth}
\centering
\includegraphics[width=1.0\textwidth]{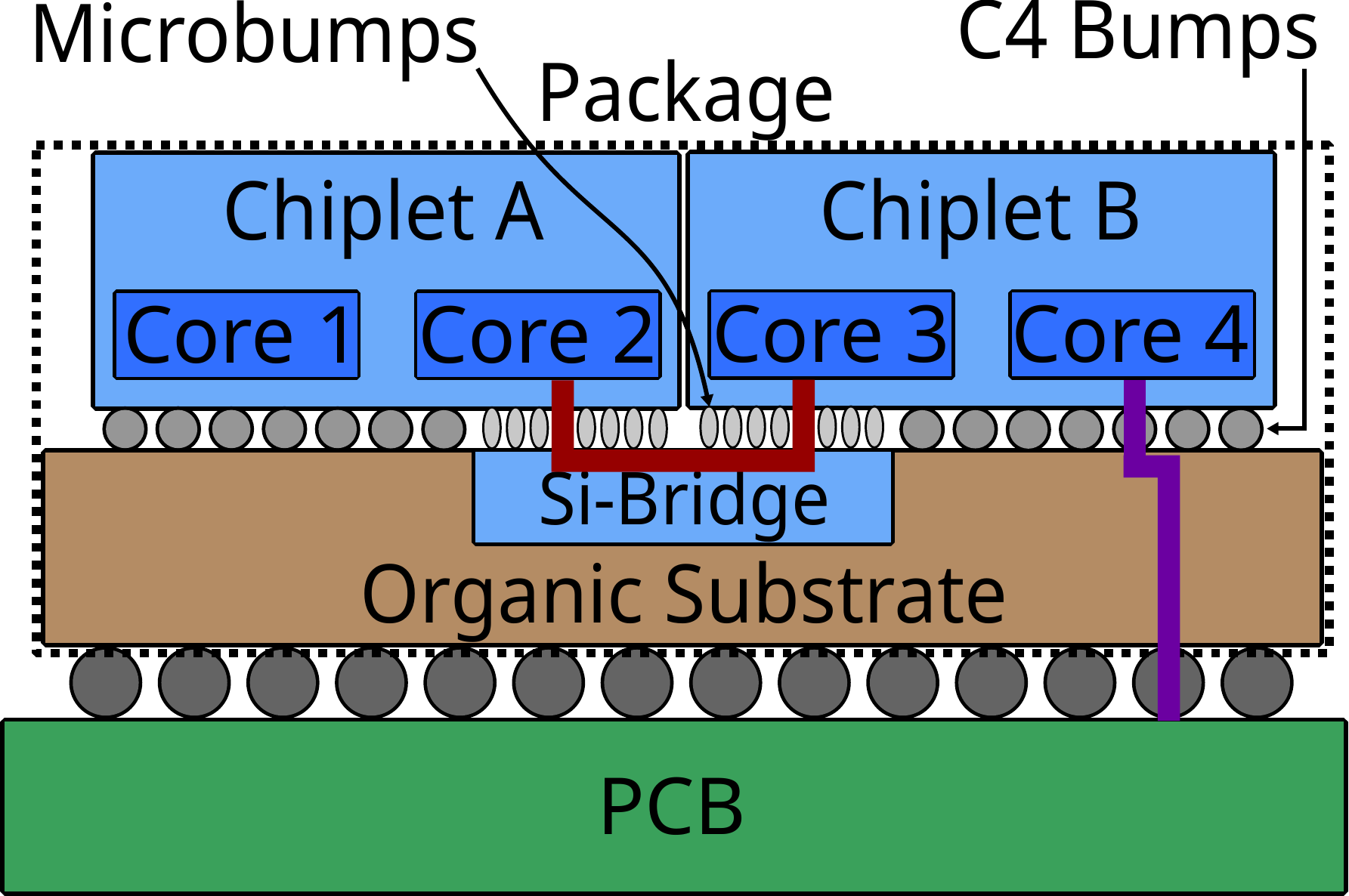}
\caption{Silicon bridge.}
\label{fig:back-bridge}
\end{subfigure}
\begin{subfigure}{0.3 \textwidth}
\centering
\includegraphics[width=1.0\textwidth]{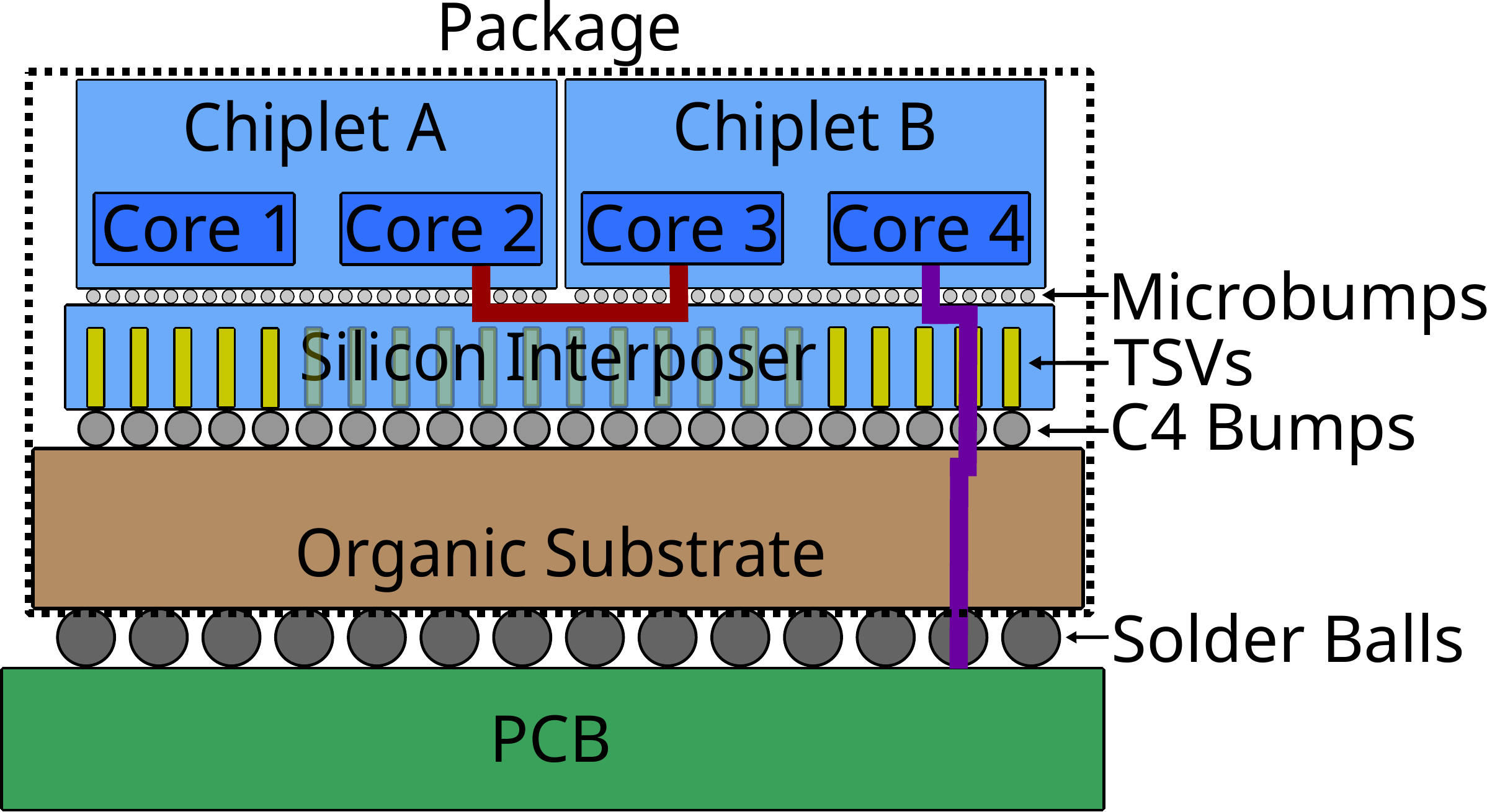}
\caption{Silicon interposer (active or passive)}
\label{fig:back-interposer}
\end{subfigure}
\caption{\textbf{(\textsection \ref{ssec:back-25D}) 2.5D integration technologies} (side view). We show a core-to-core link (red) and an off-chip link (purple).}
\label{fig:back-integration}
\end{figure*}
\section{Introduction}
\label{sec:intro}

\ps{Challenges of technology scaling}

The growing demand for computing performance has always been met by increasing the number of transistors per chip, which is only possible due to CMOS technology scaling.
However, as we keep pushing the boundaries of technology scaling, we encounter multiple challenges.
Firstly, whenever we transition to a more advanced technology node, the non-recurring cost due to physical design, verification, software, mask sets, and prototyping almost doubles \cite{cost-tech-node}.
As a result, designing a chip in an advanced technology node is only economically viable if the chip is manufactured in vast quantities.
Secondly, many chip components such as I/O drivers, analog circuits, or \gls{srams} have reached their scaling limit.
This means that we cannot shrink these components further, even if we use a more advanced technology with a smaller feature size.
Thirdly, advanced technology nodes suffer from high defect rates, diminishing the yield and inflating the recurring cost.
To tackle these challenges, new chip-design paradigms have been developed.

\ps{Why 2.5D integration?}

One of these new paradigms is 2.5D integration, where multiple silicon dies called chiplets are integrated into the same package.
Once designed, a single chiplet can be reused in multiple 2.5D stacked chips, which increases the ratio of production volume to non-recurring cost.
Another advantage is that multiple chiplets - fabricated in different technologies - can be integrated into the same package.
This means that only components that can take full advantage of technology scaling are built in bleeding-edge technologies.
Components that have reached their scaling limit are fabricated in more mature and hence less costly technology nodes.
Furthermore, chiplets are smaller than monolithic chips.
Therefore, manufacturing chiplets results in less silicon area loss due to fabrication defects and hence a higher yield.
Due to these economic advantages, chip vendors such as AMD \cite{amd-chiplet} and NVIDIA \cite{chiplet-book} have adopted the 2.5D integration paradigm.  

\ps{Challenges of 2.5D integration}

An important challenge when designing 2.5D stacked chips is the construction of a low-latency and high-throughput \gls{ici}. 
To build an \gls{ici}, we connect different chiplets using \gls{d2d} links.
These links are fabricated in an organic package substrate, silicon bridge, or silicon interposer, and they are connected to the chiplets using \gls{c4} bumps or microbumps.
The number of bumps per chiplet is limited, and so is the bandwidth of \gls{d2d} links.
In addition to having lower bandwidth than links in monolithic chips, \gls{d2d} links also have higher latency.
This latency is caused by wire delay and by \gls{phys} that are necessary in both the sending and the receiving chiplet.
\gls{phys} are needed to convert between protocols, voltage levels, and frequencies, which are usually different between on-chiplet links and \gls{d2d} links.
Due to these limitations, the \gls{ici} can quickly become a bottleneck.

\ps{How we solve these challenges differently than the related work does.}

Existing approaches to maximize the performance of the \gls{ici} either optimize the placement of chiplets (with potentially heterogeneous shapes) for a predetermined \gls{ici} topology 
\cite{ho,liu,seemuth,eris,osmolovskyi,tap25d,chiou}, select one topology out of a set of candidates \cite{coskun-1, coskun-2}, or they optimize the \gls{ici} topology for a 2D grid of homogeneously shaped chiplets on an active interposer \cite{butterdonut, cluscross, kite}.
To the best of our knowledge, there is no prior work on \gls{ici} topologies for chips with heterogeneously shaped chiplets or with passive silicon interposers or silicon bridges.
To fill this gap, we propose \name, a novel optimization methodology to jointly optimize the chiplet placement and \gls{ici} topology of such architectures.
\ifnb
\else
\newpage
\fi

\ps{Details on \name~and the key idea}

The key idea is as follows: 
We optimize the chiplet placement without a predetermined topology.
For each placement generated by an optimization algorithm, we infer a placement-based \gls{ici} topology by connecting chiplets that are in close proximity in that specific placement.
We then compute the latency and throughput of this combination of placement and topology for different traffic types.
These latencies and throughputs together with the total chip area are used to compute a user-defined quality-score of the placement, which is returned to the optimization algorithm.
Based on this quality score, the algorithm can further optimize the placement.
By following this iterative process, we jointly optimize the chiplet placement and the \gls{ici} topology.

\ps{Short evaluation-summary}

We provide our open-source framework implementing the proposed placement and topology co-optimization methodology, which we evaluate using both synthetic traffic and traffic traces.
A 2D grid of chiplets with a mesh topology is used as a baseline since many proposals for 2.5D stacked chips \cite{dataflow_accel_dnn, cifher, simba, hecaton, dojo} use such an architecture.
We reduce the latency of synthetic L1-to-L2 and L2-to-memory traffic, the two most important traffic types for cache coherency traffic, by up to 28\% and 62\% respectively.
For real traffic traces, we reduce the average packet latency for almost all traces and architectures considered (reduced by an 8\% or 18\% on average depending on the configuration of \gls{phys} within a chiplet).

\section{Background}
\label{sec:back}

\subsection{2.5D Integration}
\label{ssec:back-25D}

\ps{Introduce types of 2.5D stacked chips}

The most prominent difference between monolithic chips (\Cref{fig:back-monolithic}) and 2.5D stacked chips (\Cref{fig:back-substrate,fig:back-bridge,fig:back-interposer}) is that the former only contain a single silicon die while the later contain multiple dies which are called chiplets.
We categorize 2.5D stacked chips into those that only use an organic substrate (\Cref{fig:back-substrate}), those that use silicon bridges (\Cref{fig:back-bridge}), and those that use a silicon interposer (\Cref{fig:back-interposer}).
There are passive silicon interposers \cite{psi, ops-vs-psi} only containing metal layers fabricated in the \gls{beol} and active silicon interposers \cite{intact, first-act-int} also containing transistors fabricated in the \gls{feol}.
\Cref{tab:back-chips} summarizes the packaging options.

\begin{table}[h]
\setlength{\tabcolsep}{3pt}
\centering
\captionsetup{justification=centering}
\begin{tabular}{lcccc}
\toprule
Technology & 
Cost & 
\makecell{Link\\Bandwidth} & 
\makecell{Maximum\\Link Length$^*$} &
\makecell{Package-Level\\Routers}
\vspace{-0.3em}\\
\midrule
Monolithic Chip		& High	& High	& unlimited						& n/a	\\
Organic Substrate 	& Low	& Low 	& 10-25mm$^{**}$ \cite{bow-spec}& No 	\\
Silicon Bridge		& Mid	& Mid	& 2-4mm	\cite{emib, sib-2}		& No 	\\
Passive Interposer 	& Mid	& Mid	& 2-4mm	\cite{bow-spec}			& No 	\\
Active Interposer 	& High	& Mid	& unlimited						& Yes	\\
\bottomrule
\end{tabular}
\vspace{-0.5em}
\caption{\textbf{(\textsection\ref{ssec:back-25D}) Overview of packaging technologies.} *Maximum link length depends on data rate, bump pitch, etc. **Source termination (up to 50 mm for double termination).}
\vspace{-2.5em}
\label{tab:back-chips}
\end{table}

\ps{Properties and challenges of organic substrate}

\subsubsection{2.5D Stacked Chips using an Organic Substrate}
\label{sssec:back-substrate}

In 2.5D stacked chips without a silicon bridge or interposer (Fig.  \ref{fig:back-substrate}), the chiplets are directly connected to the organic substrate using \gls{c4} bumps. 
The rather large bump-pitch of $150$-$200$ $\mu$m severely limits the number of bumps that can be used to connect a chiplet to the substrate.
Therefore, only a limited number of wires is available for the construction of \gls{d2d} links. 
Consequently, wide, high-bandwidth links that are commonly used in monolithic chips cannot be used to connect different chiplets.
Instead, the data is serialized and transmitted over narrow \gls{d2d} links, which can quickly become a bottleneck.

\ps{Properties and challenges of silicon bridges}

\subsubsection{2.5D Stacked Chips using a Silicon Bridge}
\label{sssec:back-bridge}

To counteract the bandwidth bottlenecks formed by \gls{d2d} links in organic substrates, silicon bridges (Fig. \ref{fig:back-bridge}) such as Intel's \gls{emib} \cite{emib} or IBM's \gls{dbhi} \cite{dbhi} were introduced.
For off-chip communication, chiplets are connected to the package substrate using \gls{c4} bumps.
For high-bandwidth die-to-die communication, chiplets are connected to the silicon bridge using microbumps.
Since these microbumps have a pitch of $30$-$60$ $\mu$m, the number of wires and hence the bandwidth of \gls{d2d} links is about $10\times$ higher than in chips with a package substrate only.
However, these \gls{d2d} links still deliver lower bandwidth than on-die links in monolithic chips.
Furthermore, silicon bridges limit the link length \cite{emib, sib-2}.

\ps{Properties and challenges of passive interposer}

\subsubsection{2.5D Stacked Chips using a Passive Silicon Interposer}
\label{sssec:back-interposer-passive}

Another way of improving the bandwidth of \gls{d2d} links are passive silicon interposers, such as TSMC's \gls{cowos} \cite{cowos}.
Here, chiplets are connected to the interposer using microbumps and the interposer is connected to the package substrate using \gls{c4} bumps.
\gls{Tsvs} are used to provide connectivity between chiplets and the substrate.
One major limitation of passive interposers is the limited length of \gls{d2d} links.
Since passive interposers do not contain any transistors, it is not possible to build buffered links.
Therefore, unbuffered links are used, and their length cannot exceed some millimeters \cite{ucie, bow, bow-spec}.

\ps{Properties and challenges of active interposer}

\subsubsection{2.5D Stacked Chips using an Active Silicon Interposer}
\label{sssec:back-interposer-active}

Active interposers \cite{intact, first-act-int} allow the construction of buffered \gls{d2d} links, therefore, links in active interposers can be arbitrarily long.
Another advantage of active interposers is that they allow the integration of package-level routers.
However, active interposers come with their own challenges.
Since active interposers require the \gls{feol} process, they are about ten times more expensive than passive interposers \cite{coskun-2}.
Furthermore, they suffer from a lower yield than passive interposers, which is problematic as interposers are usually large.
Another challenge is the fact that active interposers generate heat, which is hard to remove since the interposer sits below the chiplets.
Due to these drawbacks of active silicon interposers, our work focuses on the remaining 2.5D integration technologies.

\subsection{Optimization Algorithms Suitable for Chiplet Placements}
\label{ssec:back-opt}

\ps{Quick intro of best random}

\subsubsection{Best Random}
\label{sssec:back-opt-br}

This algorithm generates random solutions to an optimization problem until the time budget is exhausted, then, it returns the best solution that was found.
We use this na\"ive algorithm as a baseline to determine whether the more advanced algorithms perform better than random.

\ps{Quick intro of the genetic algorithm}

\subsubsection{The Genetic Algorithm}
\label{sssec:back-opt-ga}

The \gls{ga} mimics biological evolution, where a solution to an optimization problem is viewed as an individual.
Each individual has a fitness score, corresponding to the quality of the solution.
The algorithm maintains a population of $P$ individuals, which changes through a series of generations.
In each generation, individuals with an insufficient fitness score are eliminated from the population and existing individuals with good fitness scores are merged to produce new individuals.
The key idea is to merge two solutions with good properties to achieve a new solution with even better properties.
For the genetic algorithm to work properly, we need to formulate a merge function that enables combining the strengths of two solutions while eliminating their weaknesses.

\ps{Quick intro of simulated annealing}

\subsubsection{Simulated Annealing}
\label{sssec:back-opt-sa}

The \gls{sa} algorithm is based on the annealing process, in which a material is heated up and slowly cooled down in order to alter its properties.
We start with a randomly selected solution and alter this solution through a series of iterations.
In each iteration, we explore a \textit{neighboring} solution.
If this solution is better than the current one, we accept it, i.e., we set it as our current solution.
Otherwise, we only accept it with a certain probability. 
This probability decreases over time since we assume that the longer our algorithm is running, the closer to the optimum we are and hence the less favorable it is to accept an inferior solution.
When formulating a problem for \gls{sa}, we need a method to generate \textit{neighboring} solutions with a similar quality.
If there is no sufficient correlation between the quality of a solution and the quality of its neighbors, then simulated annealing can not work properly.

\section{Placement and Topology Co-Optimization}
\label{sec:coopt}

\ps{How performance depends on combination of topology and placement}

2.5D stacked chips require low-latency and high-throughput \gls{ici}s.
The latency mainly depends on the number of chiplet-to-chiplet hops and on the link latency.
The throughput primarily depends on the frequency, at which the links can be operated and on the congestion, i.e. how many different flows compete for the same link.
While hop count and congestion both depend on the \gls{ici} topology, link latency and link frequency both depend on the link length, which depends on the combination of \gls{ici} topology and chiplet placement.
To maximize the performance of the \gls{ici}, chiplet placement and \gls{ici} topology must be perfectly aligned.
Selecting a topology first and then optimizing the placement for that topology might not yield satisfying results, since the choice of a suboptimal topology can prevent us from finding a good placement.
Selecting a placement first and the optimizing the topology results in a similar problem.
The solution to this problem is to co-optimize the chiplet placement and \gls{ici} topology.

\begin{figure}[H]
\vspace{-1em}
\centering
\captionsetup{justification=centering}
\includegraphics[width=1.0\columnwidth]{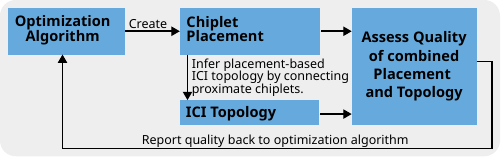}
\caption{\textbf{(\textsection\ref{sec:coopt}) Placement and topology co-optimization.}}
\label{fig:coopt-idea}
\vspace{-0.5em}
\end{figure}

\ps{Explain PlaceIT}

\Cref{fig:coopt-idea} visualizes our proposed chiplet placement and \gls{ici} topology co-optimization methodology.
An optimization algorithm is used to optimize the chiplet placement.
For each placement that the optimization algorithm produces, a placement-based \gls{ici} topology is inferred.
In this inference process, we minimize the length of \gls{d2d} links.
We then assess the quality of the combined chiplet placement and \gls{ici} topology, which we return to the optimization algorithm.
Using this methodology, the optimization algorithm does not only optimize metrics of the placement itself, like, e.g., the total area, but it also optimizes the placement in a way that enables us to construct good \gls{ici} topologies on top of it.
This is comparable to placers in the \gls{vlsi} place \& route step that optimize the placement of macros not only for area but also for routability of wires.
The difference to our methodology is that placers optimize the macro placement for the routability of predetermined nets, while we optimize the chiplet placement for the inference of an \gls{ici} topology with a yet unknown connectivity pattern.

\begin{figure*}[h]
\centering
\captionsetup{justification=centering}
\includegraphics[width=0.9\textwidth]{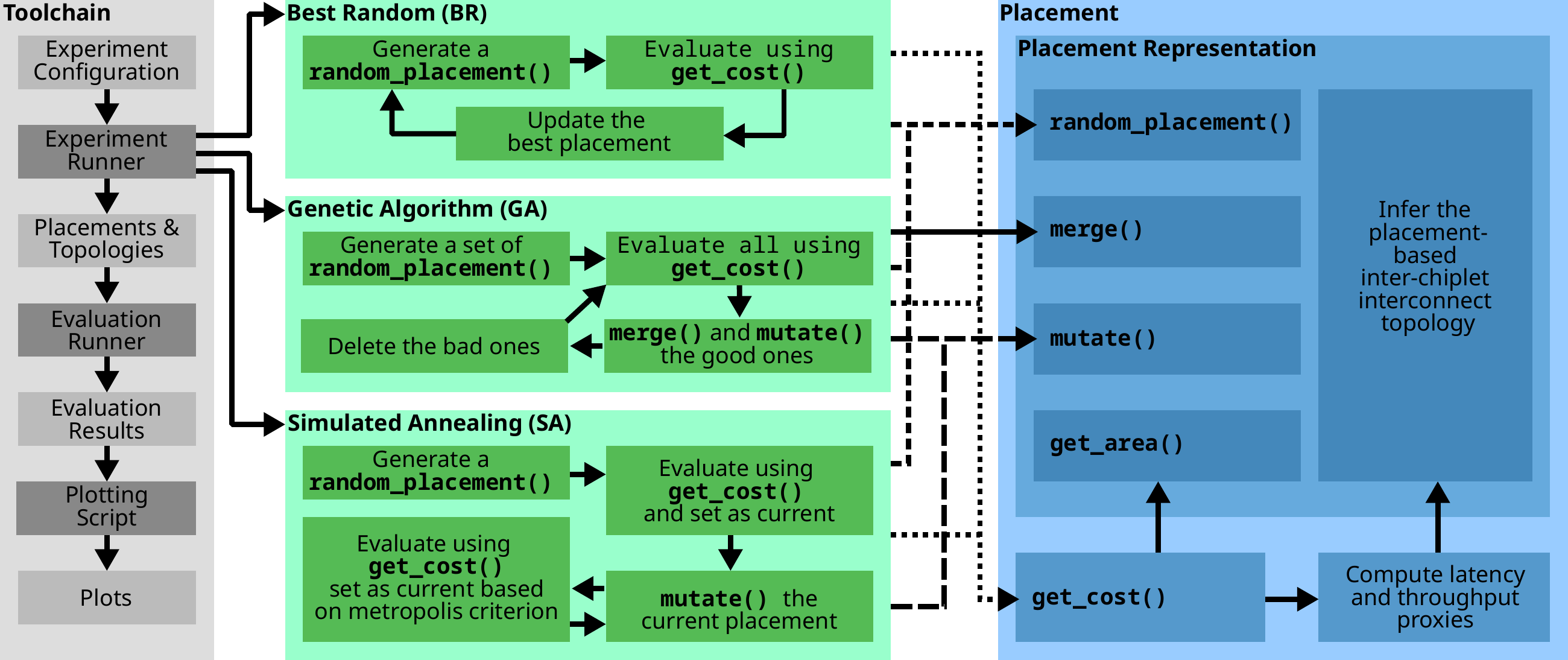}
\caption{\textbf{(\textsection \ref{sec:arch}) Overview of the \name~architecture.}}
\label{fig:arch-arch}
\vspace{-2em}
\end{figure*}
\section{\name~Architecture}
\label{sec:arch}

\ps{State our assumptions}

In this section, we present our open-source implementation of the proposed chiplet placement and \gls{ici} topology co-optimization methodology.
Our framework is based on the following list of assumptions:

\begin{itemize}
	\item Every chiplet is categorized as compute, memory, or IO.
	\item We know the number and locations of \gls{phys} in each chiplet.
	\item \gls{phys} use a common protocol, e.g., \gls{ucie} \cite{ucie} or \gls{bow} \cite{bow}.
	\item The data-width of all \gls{phys} is the same, i.e., we can connect any two \gls{phys} with a \gls{d2d} link.
	\item For each chiplet, we know whether it can relay traffic. Relaying traffic means that a message enters the chiplet through one PHY and leaves it through a different PHY.
	\item Off-chip traffic leaves the chip through an IO-chiplet as in the AMD EPYC and Ryzen processor families \cite{amd-chiplet}.
\end{itemize}

\ps{Give rough outline of the \name~architecture}

\Cref{fig:arch-arch} provides an overview of the \name~architecture. 
The experiment configuration contains the general configuration of \name~and parameters describing the architecture to be optimized (see \Cref{tab:arch-config}) as well as parameters for the three optimization algorithms.
The experiment runner launches multiple runs of each optimization algorithm.
We support the optimization algorithms \acrfull{br}, \acrfull{ga}, and \acrfull{sa}, but custom ones can be added.
Optimization algorithms interact with the placement through the following functions:

\begin{itemize}
	\item	\texttt{random\_placement()}: Return a random placement.
	\item	\texttt{mutate(X)}: Return an altered version of placement X.
	\item	\texttt{merge(X, Y)}: Return a hybrid of placements X and Y.
	\item	\texttt{get\_cost()}: Return the placement's cost.
\end{itemize}

\begin{table}[h]
\setlength{\tabcolsep}{2pt}
\centering
\captionsetup{justification=centering}
\vspace{-1em}
\begin{tabular}{ll}
\toprule
\multicolumn{2}{l}{General \name~configuration}\\
\midrule
Opt. Algorithm		& \emph{Best Random}, \emph{Genetic Algorithm}, \emph{Simulated Annealing}.		\\
Placement Repr.		& \emph{Homogeneous} or \emph{Heterogeneous}.									\\
Time Budget			& Run the algorithm for this many seconds.										\\
Repetitions			& Perform this many repetitions of the experiment.								\\
Norm. Samples		& \makecell[l]{Number of samples used to compute the cost-function				\\
				  	  normalizers (see \Cref{ssec:arch-cost}).}								\\
Mutation Mode		& How to mutate placements (see \Cref{ssec:homo-repr,ssec:hetero-repr}).		\\
\midrule
\multicolumn{2}{l}{Parameters specifying the architecture to be optimized}\\
\midrule
Distance Type		& \emph{Manhattan} or \emph{Euclidean} distance.								\\
Max. Length			& The maximum length of a \gls{d2d} link [mm].									\\
N$_{C}$,N$_{M}$,N$_{I}$ & Number of compute-, memory-, and IO-chiplets.							\\
Dimensions			& The width and height of each chiplet type [mm].								\\
PHYs				& The number and position of PHYs in each chiplet type.							\\
Relay Chiplets		& List of chiplets that can relay traffic.									\\
L$_{P}$,L$_{L}$,L$_{R}$ & \makecell[l]{Latency of PHYs, links, and relaying a message\\through a chiplet [cycles].}\\
\bottomrule
\end{tabular}
\caption{\textbf{(\textsection \ref{sec:arch}) Experiment configuration parameters.}}
\label{tab:arch-config}
\vspace{-1em}
\end{table}

The core of the placement is the placement representation, which implements the first three out of the four functions listed above.
We provide two placement representations, one for homogeneously shaped chiplets (see \Cref{sec:homo}) and one for heterogeneously shaped chiplets (see \Cref{sec:hetero}), but custom ones can be implemented.
In addition to the \texttt{random\_placement()}, \texttt{merge()}, and \texttt{mutate()} functions, the placement representation contains functions to get the area and the placement-based \gls{ici} topology.
To assess the quality of a placement, we infer the placement-based \gls{ici} topology and use it to estimate the \gls{ici} latency and throughput, which we combine with the total chip area to compute a user-defined cost function.
Our performance estimates and cost function are explained in detail in the following sections.

\subsection{Computation of Performance Estimates}
\label{ssec:arch-proxies}

We use the RapidChiplet \cite{rapidchiplet} toolchain to estimate the latency and throughput of the \gls{ici}. RapidChiplet provides high-level latency and throughput proxies for \gls{c2c}, \gls{c2m}, \gls{c2i}, and \gls{m2i} traffic as well as simulation-based results for both synthetic traffic patterns and application traces.

\ps{Explain the cost function}

\begin{figure*}[h]
\centering
\captionsetup{justification=centering}
\includegraphics[width=0.99\textwidth]{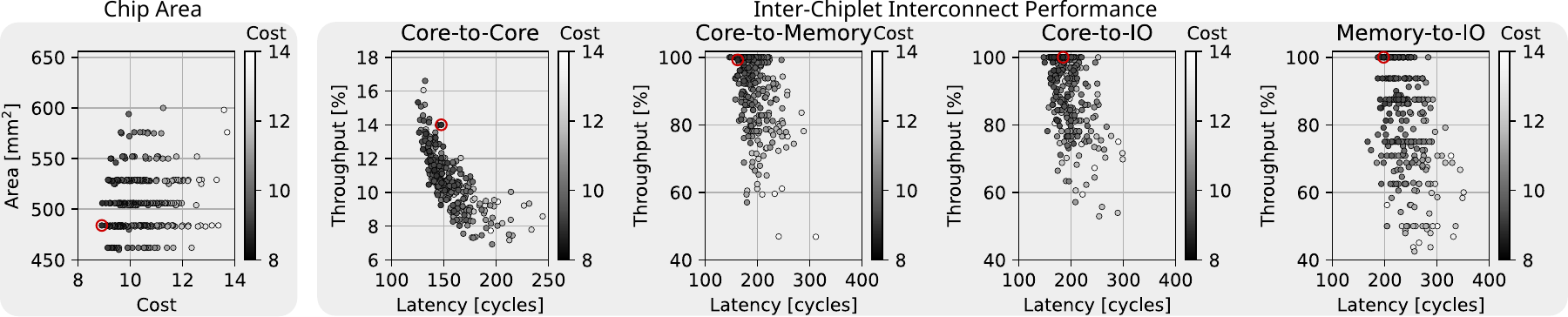}
\caption{\textbf{(\textsection \ref{ssec:arch-cost}) Correlation of cost value with its components.}
Points corresponds to random designs with colors indicating the placement's cost.
Red circles highlight the lowest-cost placement.
Throughput is given in percent of the theoretical peak.
}
\label{fig:arch-set-weight}
\vspace{-1.0em}
\end{figure*}

\subsection{The Cost Function}
\label{ssec:arch-cost}

Specifying the cost function is the most crucial part of applying the \name~framework, as it sets the goal towards which the chiplet placement and \gls{ici} topology are co-optimized.

As a running example, throughput this paper, we use \name~to build a general-purpose cache-coherent system targeted at various scientific simulation workloads from the high-performance computing domain. 
To that end, we use a weighted sum of the latency and throughput of \gls{c2c}, \gls{c2m}, \gls{c2i}, and \gls{m2i} traffic. 
In addition, we incorporate a term for the area of a minimal rectangle fully enclosing all chiplets (from now on referred to as the area) to ensure that \name~produces compact placements.
\Cref{fig:arch-set-weight} visualizes the correlation of the final cost value with its nine components (four latency and four throughput terms plus the area term).

The user-defined cost function makes \name~applicable to a wide range of optimization goals. One could for example use estimates of the \gls{ici} latency and throughput under a certain application trace or a set of application traces to design a domain-specify accelerator, e.g., for machine learning training and inference, image and video processing, or graph analytics.

\section{\name~for Homogeneous Chiplet Shapes}
\label{sec:homo}

\ps{Introduce assumptions specific to the homogeneous setting}

We discuss our placement representation and optimization results for chiplets with homogeneous shapes.
In this setting, we assume that there are two possible configurations of \gls{phys} for a chiplet:
A chiplet either has four \gls{phys}, one on each side (we use this for compute-chiplets), or a single PHY on one side (we use this for memory- and IO-chiplets).

\subsection{Placement Representation}
\label{ssec:homo-repr}

\ps{Introduce the key idea of the placement representation for homogeneous chiplets}

\begin{figure}[H]
\centering
\captionsetup{justification=centering}
\begin{subfigure}{1.0 \columnwidth}
\centering
\includegraphics[width=1.0\columnwidth]{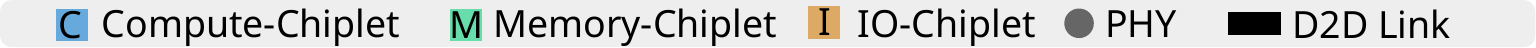}
\vspace{-1.0em}
\end{subfigure}
\begin{subfigure}{0.19 \columnwidth}
\centering
\includegraphics[width=1.0\columnwidth]{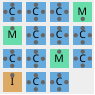}
\caption{Random placement.}
\label{fig:homo-plac-init}
\end{subfigure}
\begin{subfigure}{0.19 \columnwidth}
\centering
\includegraphics[width=1.0\columnwidth]{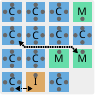}
\caption{Two mutations.}
\label{fig:homo-plac-mutate}
\end{subfigure}
\begin{subfigure}{0.19 \columnwidth}
\centering
\includegraphics[width=1.0\columnwidth]{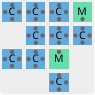}
\caption{Merge step one.}
\label{fig:homo-plac-merge-1}
\end{subfigure}
\begin{subfigure}{0.19 \columnwidth}
\centering
\includegraphics[width=1.0\columnwidth]{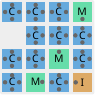}
\caption{Merge step two.}
\label{fig:homo-plac-merge-2}
\end{subfigure}
\begin{subfigure}{0.19 \columnwidth}
\centering
\includegraphics[width=1.0\columnwidth]{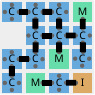}
\caption{Extract network.}
\label{fig:homo-plac-network}
\end{subfigure}
\caption{
\textbf{(\textsection \ref{ssec:homo-repr}) Homogeneous placement representation.}
\textbf{(b)} is a mutation of (a), \textbf{(c) and (d)} show the process of merging (a) and (b), \textbf{(e)} extracts the network of (d).
}
\vspace{-0.75em}
\label{fig:homo-plac}
\end{figure}

We represent a placement of homogeneously shaped chiplets as an $R \times C$ grid such that $R \cdot C \geq N_\text{C} + N_\text{M} + N_\text{I}$.
A grid cell can contain a compute-, memory-,  or IO-chiplet, or it can be empty (see \Cref{fig:homo-plac-init}).
Chiplets with a single PHY can be rotated, but chiplets with four PHYs cannot, since this would result in an isomorphic placement with identical cost.
Furthermore, chiplets with a single PHY are always placed in a way s.t. the PHY faces another chiplet and not the "outside".
We provide the following functions for homogeneous placements:

\ps{Explain details of the placement representation for homogeneous chiplets}

\begin{itemize}
	\item \texttt{random\_placement()}:
	Randomly place all chiplets in the $R \times C$ grid (see \Cref{fig:homo-plac-init} for an example).
	\item	\texttt{mutate(X)}
	We provide four different mutation modes.
	\begin{itemize}
		\item \emph{any-both}: swap two chiplets \textbf{and} rotate one.
		\item \emph{any-one}: swap two chiplets \textbf{or} rotate one.
		\item \emph{neighbor-both}: swap two adjacent chiplets \textbf{and} rotate one.
		\item \emph{neighbor-one}: swap two adjacent chiplets \textbf{or} rotate one.
	\end{itemize}
	Only chiplets of different types can be swapped, since swapping two chiplets of the same type would result in an isomorphic placement with identical cost.
	\Cref{fig:homo-plac-mutate} visualizes an \emph{any-one} and a \emph{neighbor-one} mutation.

	\item	\texttt{merge(X,Y)}
	If chiplet types and/or chiplet rotations match between placements X and Y, then, those types and/or rotations are carried over to the merged placement.
	The remaining, unplaced chiplets are placed and rotated randomly.
	\Cref{fig:homo-plac-merge-1,fig:homo-plac-merge-2} visualize the merge process.

	\item	\texttt{get\_network()}
	The placement-based \gls{ici} topology is created by connecting any two opposing \gls{phys} of adjacent chiplets (see \Cref{fig:homo-plac-network}).
	This might result in an unconnected placement (as in \Cref{fig:homo-plac-network}).
	In those cases, the operation that created the unconnected placement is repeated.

	\item	\texttt{get\_area()}
	The area of a homogeneous placement is chiplet\_size $\cdot R \cdot C$.
	Consequently, in the homogeneous case, all placements of the same architecture have the same area.
\end{itemize}

\subsection{Optimization Results}
\label{ssec:homo-opt}

\ps{Provide the experiment configuration}

We use \name~to optimize two architectures, one with $32$ compute-, $4$ memory-, and $4$ IO-chiplets (called \textit{32 cores homogeneous}) and one with $64$ compute-, $8$ memory-, and $8$ IO-chiplets (called \textit{64 cores homogeneous}).
Both architectures use chiplets of size $3$mm$ \times $3$mm$.
We perform our experiments on a single core of an Intel Xeon X7550 running Debian 11.

To set the weights of \gls{c2c}, \gls{c2m}, \gls{c2i}, and \gls{m2i} latency and throughput in the cost function accordingly, we analyze multiple cache-coherency traffic traces from the Netraces v1.0 \cite{netrace-traces} trace collection.
We observe that $0\%$-$5\%$ of the messages are \gls{c2c} traffic, $80\%$-$95\%$ are \gls{c2m} traffic, and $3\%$-$16\%$ are \gls{m2i} traffic.
Therefore, we set the cost function weights for the area as well as \gls{c2m} and \gls{m2i} latency and throughput to $2$ while we set the weights for \gls{c2c} and \gls{c2i} latency and throughput to $0.1$.
\Cref{tab:homo-params} shows the remaining parameters. 

\begin{table}[h]
\setlength{\tabcolsep}{5pt}
\vspace{-0.5em}
\centering
\captionsetup{justification=centering}
\begin{tabular}{llcc}
\toprule
&Parameter					& \makecell{32 cores\\homogeneous}	& \makecell{64 cores\\homogeneous} 	\vspace{-0.3em}\\
\midrule
\multirow{4}{*}{\rotatebox[origin = c]{90}{General~~~}}
&Time Budget				& 3600 s					& 3600 s					\\	
&Repetitions				& 10						& 10 						\\
&Norm. Samples				& 500 						& 500 						\\	
&Mutation Mode				& neighbors-one 			& neighbors-one 			\\	
&$L_\text{R}$,$L_\text{P}$, and $L_\text{L}$		& 10, 12, and 1 cycles 		& 10, 12, and 1 cycles 				\\	
\midrule
\multirow{4}{*}{\rotatebox[origin = c]{90}{GA}}
&Population size (P)		& 200  						& 50 						\\	
&Elitism size (E)			& 30  						& 8 						\\	
&Tournament size (T)		& 30  						& 8 						\\	
&Mutation prob. ($p_m$) 	& 0.5						& 0.5 						\\	
\midrule
\multirow{4}{*}{\rotatebox[origin = c]{90}{SA}}
&Initial Temp. ($T_0$) 		& 40 						& 35 						\\	
&Iterations ($L$) 			& 250						& 50 						\\	
&Cooling param. ($\alpha$) 	& 1 						& 1 						\\	
&Adaptive param. ($\beta$) 	& 5 						& 5 						\\	
\bottomrule
\end{tabular}
\caption{\textbf{(\textsection \ref{ssec:homo-opt}) Experiment configuration}\\for homogeneously shaped chiplets.}
\label{tab:homo-params}
\end{table}

\ps{Direct comparison between algorithms}

\begin{figure}[h]
\centering
\captionsetup{justification=centering}
\vspace{-1em}
\includegraphics[width=1.0\columnwidth]{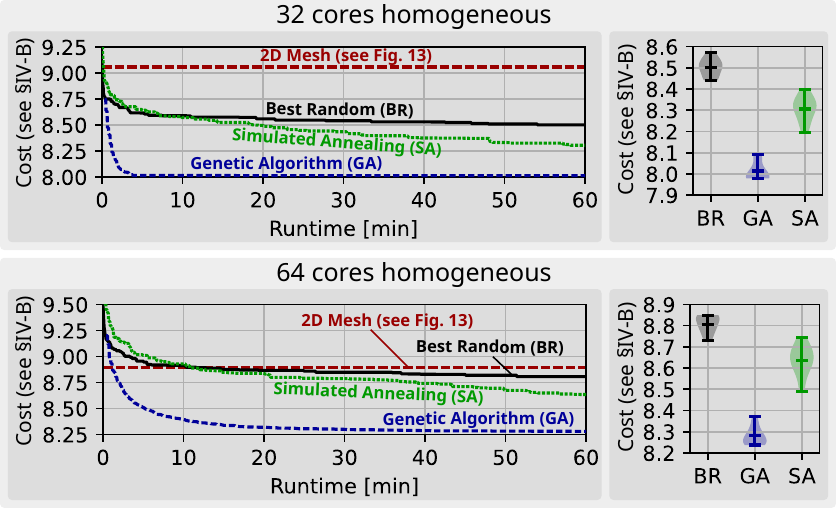}
\caption{\textbf{(\textsection \ref{ssec:homo-opt}) Results for homogeneously shaped chiplets.}
We show the evolution of the result over time (left) and the distribution of the final result over 10 repetitions (right).
See \Cref{fig:eval-placements} for the placements found by the best algorithm.}
\vspace{-0.75em}
\label{fig:homo-results}
\end{figure}

\Cref{fig:homo-results} shows our results for the two architectures.
For both architectures, all three optimization algorithms are able to outperform the baseline architecture which is based on a 2D mesh (see \Cref{fig:eval-placements}).
This is not surprising, as such an architecture is not ideal for \gls{c2m} and \gls{m2i} traffic, which was the highest weighted traffic in our cost function.
Furthermore, we observe that both the \gls{ga} and \gls{sa} significantly outperform \gls{br} which shows that the two more complex optimization algorithms work as intended. 

\ps{Discuss convergence}

We observe that the \gls{ga} is able to quickly converge to a good solution. 
For the 32 core architecture with a solution space of approximately $10^{14}$ solutions, the \gls{ga} converges after five minutes.
For the 64 core architecture with a solution space of approximately $10^{30}$ solutions, the convergence takes about 30 minutes.
\gls{sa} is able to continuously increase the quality of the solution, however, the allocated compute budget was not sufficient for \gls{sa} to reach convergence.
This shows the advantage of maintaining multiple good solutions and combining them (\gls{ga}) over keeping a single solution (\gls{sa}).

\section{\name~for Heterogeneous Chiplet Shapes}
\label{sec:hetero}

\ps{Section introduction}

We explain our method to represent and optimize placements of heterogeneously shaped chiplets with arbitrary, rectangular shapes and arbitrary PHY counts and positions.

\subsection{Placement Representation}
\label{ssec:hetero-repr}

\ps{Introduce the first idea one might have and why we don't use it}

An intuitive representation of a placement of heterogeneously shaped chiplets would be a list of chiplets where each chiplet has a location and a rotation.
However, with such an approach, the \texttt{mutate()} and \texttt{merge()} operations could result in overlapping chiplets.
Only allowing operations that do not result in overlaps would severely restrict the number of operations and prohibit an exploration of the full solution space.
Allowing operations that result in overlaps and using a legalization step to make the chiplets non-overlapping again makes the placement quality deteriorate over time (we tried this as our first approach).
Therefore, in the remainder of this section, we present our more elaborate placement representation for chiplets with heterogeneous shapes.

\ps{Introduce the key idea of our representation of heterogeneous chiplets}

\begin{figure}[H]
\centering
\vspace{-1em}
\captionsetup{justification=centering}
\begin{subfigure}{1.0 \columnwidth}
\centering
\includegraphics[width=1.0\columnwidth]{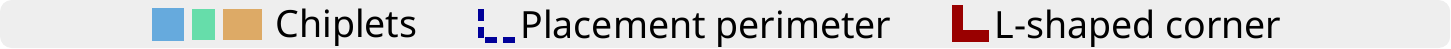}
\vspace{-1.0em}
\end{subfigure}
\begin{subfigure}{0.24 \columnwidth}
\centering
\includegraphics[width=0.95\columnwidth]{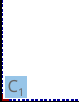}
\caption{Placing $C_1$.\\~}
\label{fig:hetero-plac-1}
\end{subfigure}
\begin{subfigure}{0.24 \columnwidth}
\centering
\includegraphics[width=0.95\columnwidth]{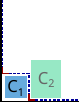}
\caption{Placing $C_2$.\\~}
\label{fig:hetero-plac-2}
\end{subfigure}
\begin{subfigure}{0.24 \columnwidth}
\centering
\includegraphics[width=0.95\columnwidth]{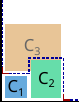}
\caption{Placing $C_3$\\creates overlap.}
\label{fig:hetero-plac-3}
\end{subfigure}
\begin{subfigure}{0.24 \columnwidth}
\centering
\includegraphics[width=0.95\columnwidth]{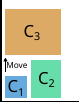}
\caption{Move $C_3$ to\\fix overlap.}
\label{fig:hetero-plac-4}
\end{subfigure}
\vspace{-2em}
\caption{\textbf{(\textsection \ref{ssec:hetero-repr}) Our custom placement algorithm.}}
\label{fig:hetero-plac}
\vspace{-1em}
\end{figure}

Instead of optimizing the location of chiplets directly, we optimize the \emph{order} and \emph{rotations} in which our custom placement algorithm places the chiplets.
Like this, every possible combination of \emph{order} and \emph{rotations} results in a placement with non-overlapping chiplets.
Our custom chiplet placement algorithm iterates through the chiplets in the specified order and places each of them by performing the following steps:

\begin{enumerate}[leftmargin=*, label={Step \arabic*:}]
	\item	Draw a line along the perimeter of the current placement (see the blue, dashed line in \Cref{fig:hetero-plac}).			
	\item	Identify all L-shaped corners of the perimeter-line (see the red solid corners in \Cref{fig:hetero-plac}).
	\item	Place the chiplet in the corner, that minimizes the area of a minimum square enclosing the whole placement.
	\item	The third step can result in overlaps (see \Cref{fig:hetero-plac-3}).
			If the newly placed chiplet has an overlap on the right, move it to the top and vice versa (see \Cref{fig:hetero-plac-4}).
\end{enumerate}

\ps{Explain additional issues with our approach and how we solve them}

\begin{figure}[H]
\centering
\vspace{-0.5em}
\captionsetup{justification=centering}
\includegraphics[width=1.0\columnwidth]{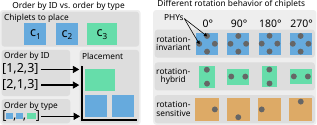}
\caption{\textbf{(\textsection \ref{ssec:hetero-repr}) Prevent multiple (order, rotations)-pairs from producing the same placement} by using order by type and dissallowing rotations based on the rotation behavior.}
\label{fig:hetero-issues}
\end{figure}

Notice that the first-class citizen on which the optimization algorithms operate are the \emph{order} and \emph{rotations} in which the chiplets are placed, not the placement itself.
Due to this indirection, we need to avoid isomorphic representations (different (order, rotations)-pairs that result in the same placement). 
If this is not the case, an optimization algorithm might use a set of supposedly different solutions that all result in the same placement.
To avoid isomorphic representations, we represent the order by chiplet types and not by chiplet IDs since two different orders by ID can result in equivalent placements, but two different orders by type cannot (see \Cref{fig:hetero-issues} left).
Furthermore, notice that a chiplet can be rotation-invariant, rotation-sensitive, or rotation-hybrid depending on whether the chiplet shape and PHY locations change upon rotation (see \Cref{fig:hetero-issues} right).
To prevent isomorphic representations, we disallow rotations of rotation-invariant chiplets, and we only allow $90^\circ$ rotations for rotation-hybrid chiplets.

\vspace{-1em}
\begin{figure*}[h]
\centering
\captionsetup{justification=centering}
\begin{subfigure}{0.94 \textwidth}
\centering
\includegraphics[width=1.0\textwidth]{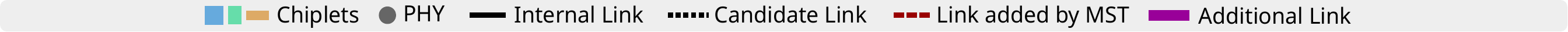}
\vspace{-1.0em}
\end{subfigure}
\\
\begin{subfigure}{0.16 \textwidth}
\centering
\includegraphics[width=0.9\textwidth]{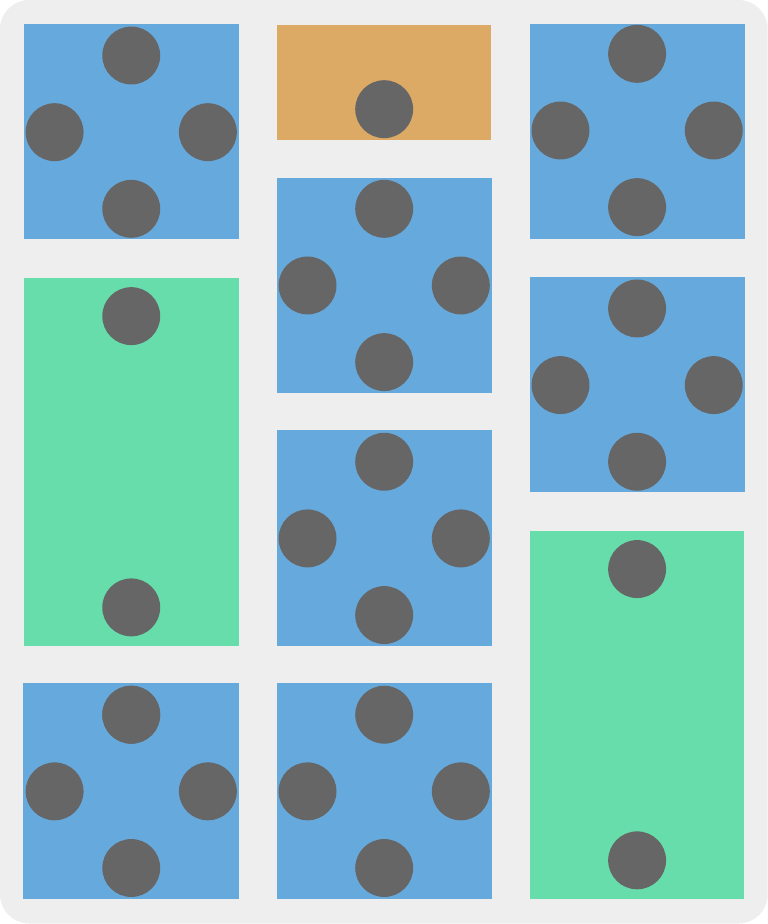}
\caption{Input: Placemet.}
\label{fig:hetero-nw-1}
\end{subfigure}
\begin{subfigure}{0.025\textwidth}
\centering
\includegraphics[width=0.9\textwidth]{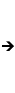}
\vspace{0.9em}
\end{subfigure}
\begin{subfigure}{0.16 \textwidth}
\centering
\includegraphics[width=0.9\textwidth]{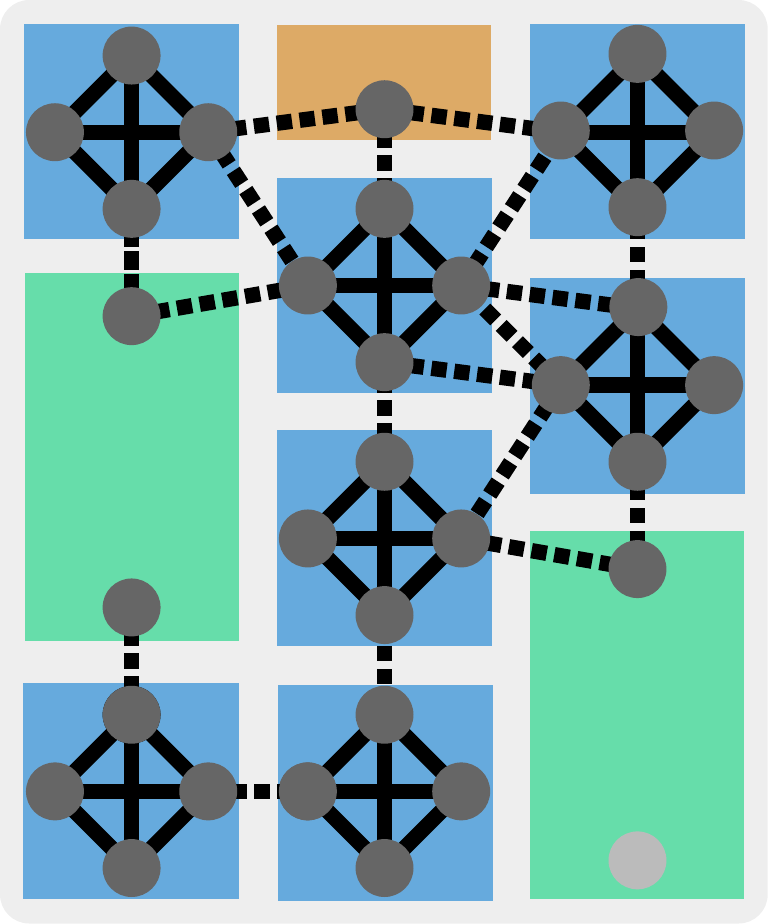}
\caption{Create graph.}
\label{fig:hetero-nw-2}
\end{subfigure}
\begin{subfigure}{0.025\textwidth}
\centering
\includegraphics[width=0.9\textwidth]{img/heterogeneous/hetero_nw_arrow.pdf}
\vspace{0.9em}
\end{subfigure}
\begin{subfigure}{0.16 \textwidth}
\centering
\includegraphics[width=0.9\textwidth]{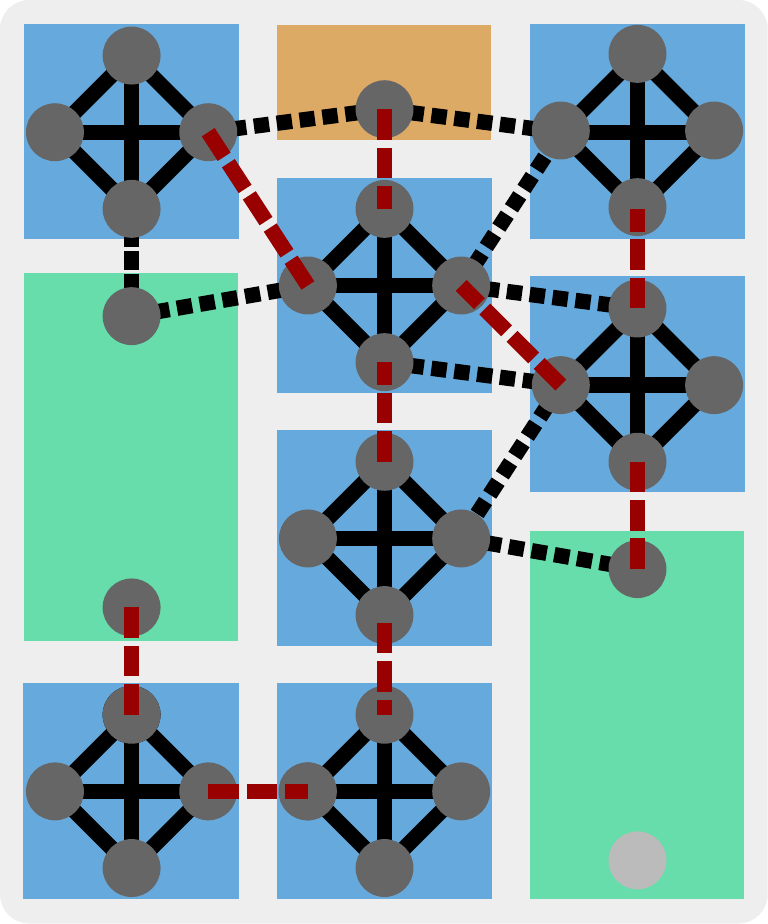}
\caption{Compute MST.}
\label{fig:hetero-nw-3}
\end{subfigure}
\begin{subfigure}{0.025\textwidth}
\centering
\includegraphics[width=0.9\textwidth]{img/heterogeneous/hetero_nw_arrow.pdf}
\vspace{0.9em}
\end{subfigure}
\begin{subfigure}{0.16 \textwidth}
\centering
\includegraphics[width=0.9\textwidth]{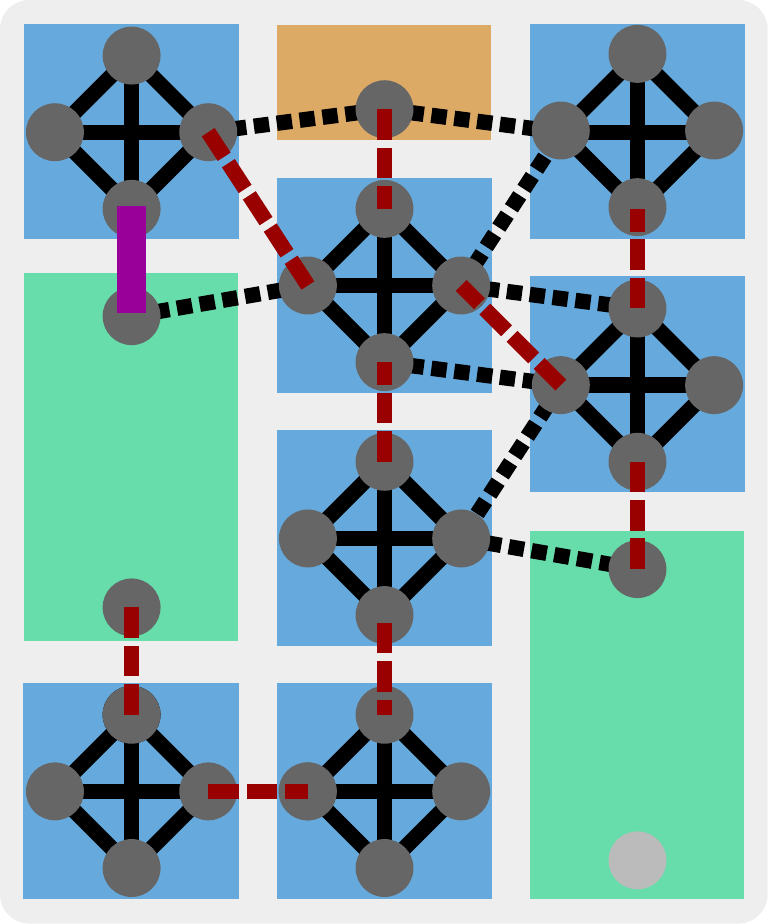}
\caption{Add more edges.}
\label{fig:hetero-nw-4}
\end{subfigure}
\begin{subfigure}{0.025\textwidth}
\centering
\includegraphics[width=0.9\textwidth]{img/heterogeneous/hetero_nw_arrow.pdf}
\vspace{0.9em}
\end{subfigure}
\begin{subfigure}{0.16 \textwidth}
\centering
\includegraphics[width=0.9\textwidth]{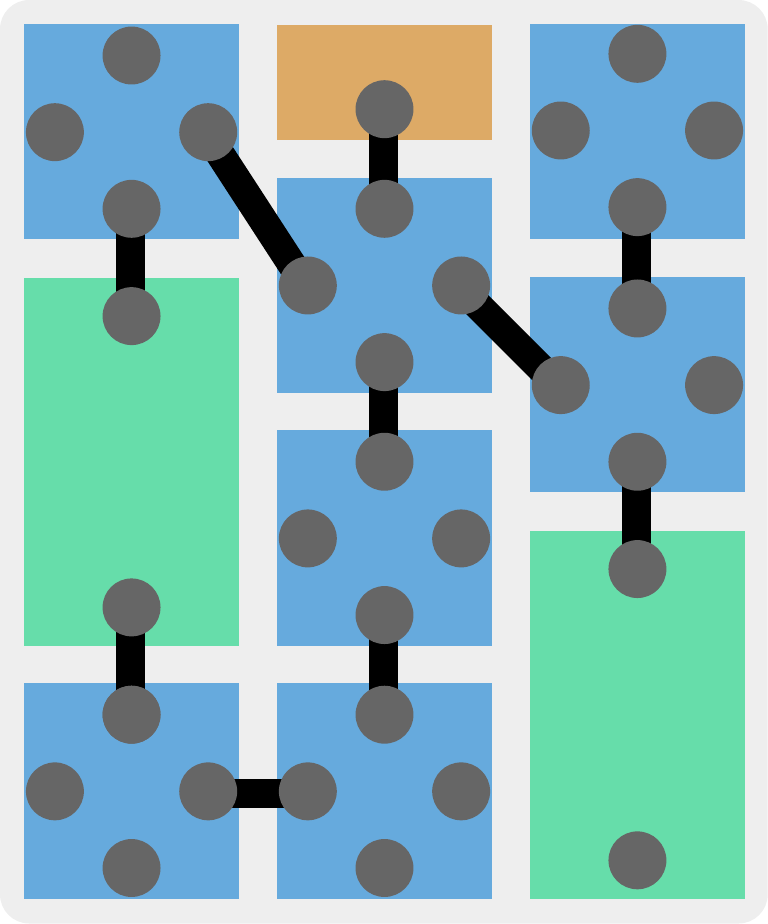}
\caption{Output: Topology.}
\label{fig:hetero-nw-5}
\end{subfigure}
\caption{\textbf{(\textsection \ref{ssec:hetero-repr}) Inferring the placement-based \gls{ici} topology} from our representation for heterogeneous chiplet placements.}
\label{fig:hetero-nw}
\end{figure*}

\ps{Explain the details of our representation of heterogeneous chiplets}

\begin{figure}[h]
\centering
\captionsetup{justification=centering}
\begin{subfigure}{0.99 \columnwidth}
\centering
\includegraphics[width=1.0\columnwidth]{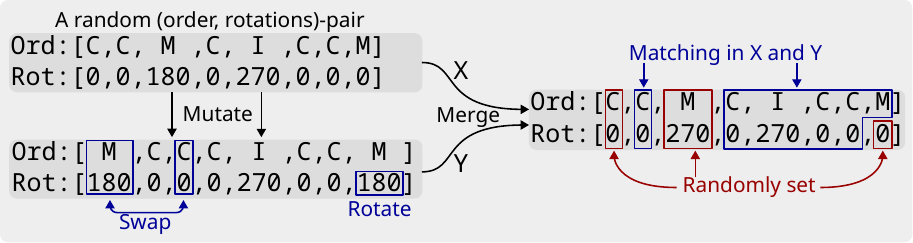}
\end{subfigure}
\caption{\textbf{(\textsection \ref{ssec:hetero-repr}) Mutate \& merge for  heterogeneous chip- lets.} C: Compute-chiplet, M: Memory-chiplet, I: IO-chiplet.}
\label{fig:hetero-mutate-merge}
\end{figure}

Our placement representation for heterogeneous chiplets implements the same five functions as the representation of homogeneous chiplets.
Instead of generating, mutating or merging placements directly, we perform these operations on the chiplet \emph{order} and \emph{rotations} (see \Cref{fig:hetero-mutate-merge}).
To implement the \texttt{get\_area()} function, we compute the area of a minimal rectangle that encloses the whole placement.

Extracting the placement-based \gls{ici} topology from a placement with heterogeneously shaped chiplets is more involved than for the homogeneous case: 
Given a placement (see \Cref{fig:hetero-nw-1}), we create a graph where each vertex represents a PHY.	
We add "internal" edges between all \gls{phys} of a chiplet with relay-capabilities (the black, solid edges in \Cref{fig:hetero-nw-2}).
Furthermore, we add a "candidate" edge between any two PHYs of different chiplets that are at most the maximum link length apart (the black, dotted edges in \Cref{fig:hetero-nw-2}).
We set the weight of "candidate" edges based on the link length and compute a \gls{mst} \cite{mst} on top of this graph (the red, dashed edges in \Cref{fig:hetero-nw-3}).
Each edge in the \gls{mst} corresponds to a \gls{d2d} link.
If some \gls{phys} of otherwise connected chiplets remain unconnected, we ignore them.	
If a whole chiplet remains unconnected, we abort and the operation that created this placement (random generation, mutate, or merge) is repeated.
Next, we iterate through the remaining candidate-edges, ordered by increasing weight.
If an edge is connected to two otherwise unused \gls{phys}, we add that edge to the \gls{ici} topology (the purple, fat edges in \Cref{fig:hetero-nw-4}).
\Cref{fig:hetero-nw-5} shows the resulting 	\gls{ici} topology.

\subsection{Optimization Results}
\label{ssec:hetero-opt}

\ps{Explain the Experiment Configuration}

\begin{figure}[H]
\centering
\captionsetup{justification=centering}
\begin{subfigure}{0.99 \columnwidth}
\centering
\includegraphics[width=1.0\columnwidth]{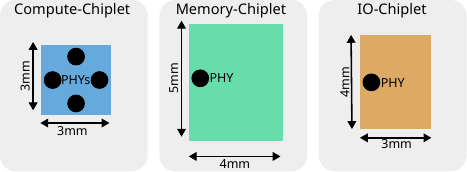}
\end{subfigure}
\caption{\textbf{(\textsection \ref{ssec:hetero-opt}) Chiplet dimensions and PHY locations.}}
\label{fig:hetero-chiplets}
\vspace{-1em}
\end{figure}

We run \name~on one core of an Intel Xeon X7550 running Debian 11 to optimize the same two architectures as in \Cref{ssec:homo-opt}, but with heterogeneous chiplets (see \Cref{fig:hetero-chiplets}).

Since we still want to optimize the \gls{ici} topology for cache coherency traffic, we use the same cost function weights as in \Cref{ssec:homo-opt}.
The remaining parameters are shown in \Cref{tab:hetero-params}.

\begin{table}[H]
\setlength{\tabcolsep}{3pt}
\centering
\captionsetup{justification=centering}
\vspace{-0.5em}
\begin{tabular}{llcc}
\hline
&Parameter					& 32 cores heterogeneous 	& 64 cores heterogeneous	\vspace{-0.3em}\\
\midrule
\multirow{4}{*}{\rotatebox[origin = c]{90}{General\qquad}}
&Time Budget				& 3600 s					& 3600 s					\\	
&Repetitions				& 10						& 10 						\\
&Norm. Samples				& 500 						& 500 						\\	
&Mutation Mode				& any-one					& any-one					\\	
&$L_\text{R}$,$L_\text{P}$, and $L_\text{L}$		& 10, 12, and 1 cycles 		& 10, 12, and 1 cycles 				\\	
&Distance Type				& Eucledian					& Eucledian					\\
&Max. Length				& 3mm						& 3mm						\\
\midrule
\multirow{4}{*}{\rotatebox[origin = c]{90}{GA}}
&Population size (P)			& 30  						& 20 						\\	
&Elitism size (E)			& 6  						& 5							\\	
&Tournament size (T)			& 6  						& 5 						\\	
&Mutation prob. ($p_m$) 		& 0.5						& 0.5 						\\	
\midrule
\multirow{4}{*}{\rotatebox[origin = c]{90}{SA}}
&Initial Temp. ($T_0$) 		& 33 						& 28 						\\	
&Iterations ($L$) 			& 50 						& 45 						\\	
&Cooling param. ($\alpha$) 	& 1 						& 1 						\\	
&Adaptive param. ($\beta$) 	& 5 						& 5 						\\	
\hline
\end{tabular}
\caption{\textbf{(\textsection \ref{ssec:hetero-opt}) Experiment configuration}\\for heterogeneously shaped chiplets.}
\label{tab:hetero-params}
\vspace{-1em}
\end{table}

\setcounter{table}{5}
\begin{table*}[b]
\setlength{\tabcolsep}{3pt}
\centering
\captionsetup{justification=centering}
\begin{tabular}{lccccccccccccccc}
\toprule
\vspace{-0.25em}
\multirow{2}{*}{Trace}		& \multicolumn{3}{c}{Region 1} & \multicolumn{3}{c}{Region 2} & \multicolumn{3}{c}{Region 3} & \multicolumn{3}{c}{Region 4} & \multicolumn{3}{c}{Region 5}\\
\cmidrule(lr){2-4} \cmidrule(lr){5-7} \cmidrule(lr){8-10} \cmidrule(lr){11-13} \cmidrule(lr){14-16}
\vspace{-0.25em}
						& P 	& C 	& I  		& P 	& C 	& I  		& P 	& C 	& I  		& P 	& C 	& I  		& P 	& C 	& I		\\
\midrule
blackscholes\_64c\_simsmall	& 189k 	& 5.6M 	& 0.0337	& 1.2M 	& 219M 	& 0.0056	& 4.9M 	& 75M 	& 0.0655	& 195k 	& 10M 	& 0.0019	& 129k 	& 5.7M 	& 0.0228	\\
bodytrack\_64c\_simlarge	& 189k 	& 5.6M 	& 0.0337	& 30M 	& 654M 	& 0.0453	& 355M 	& 3.9B 	& 0.0914	& 429k 	& 24M 	& 0.0176	& 161k 	& 5.7M 	& 0.0283	\\
canneal\_64c\_simmedium		& 189k 	& 5.6M 	& 0.0337	& 240M 	& 20B 	& 0.0121	& 74M 	& 300M 	& 0.2473	& 58M 	& 2.9B 	& 0.0198	& 133k 	& 5.7M 	& 0.0235	\\
dedup\_64c\_simmedium		& 189k 	& 5.6M 	& 0.0337	& 37M 	& 838M 	& 0.0201	& 379M 	& 2.6B 	& 0.1477	& 16M 	& 1.0B 	& 0.0153	& 160k 	& 5.7M 	& 0.0282	\\
ferret\_64c\_simmedium		& 189k 	& 5.6M 	& 0.0337	& 8.6M 	& 648M 	& 0.0133	& 273M 	& 7.5B 	& 0.0365	& 5.8M 	& 145M 	& 0.0402	& 220k 	& 5.7M 	& 0.0387	\\
fluidanimate\_64c\_simsmall	& 189k 	& 5.6M 	& 0.0337	& 6.8M 	& 777M 	& 0.0087	& 21M 	& 499M 	& 0.0420	& 6.1M 	& 599M 	& 0.0103	& 139k 	& 5.7M 	& 0.0245	\\
swaptions\_64c\_simlarge	& 189k 	& 5.6M 	& 0.0337	& 247k 	& 9.7M 	& 0.0254	& 310M 	& 1.7B 	& 0.1800	& 194k 	& 14M 	& 0.0141	& 113k 	& 5.7M 	& 0.0199	\\
x264\_64c\_simsmall			& 189k 	& 5.6M 	& 0.0337	& 1.8M 	& 82M 	& 0.0220	& 31M 	& 1.5B 	& 0.0212	& 102M 	& 12B 	& 0.0084	& 129k 	& 5.7M 	& 0.0227	\\
\bottomrule
\end{tabular}
\caption{\textbf{(\textsection \ref{ssec:eval-methodology}) Trace-regions used in the evaluation.} \textbf{P}: Number of packets, \textbf{C}: number of cycles, \textbf{I}: injection rate.}
\label{tab:eval-traces}
\end{table*}


\ps{Discuss Reduction of solution space size}

Even though the problem of placing heterogeneously shaped chiplets is significantly more complex than that of placing homogeneous ones, we were able the achieve a similar solution space size of approximately $10^{14}$ solutions for the 32 core architecture and $10^{30}$ for the 64 core architecture.
We achieve this by optimizing the order and rotations, in which our custom placement algorithm places the chiplets, not the chiplet placement itself.
Like this, we are able to remove many unfavorable or invalid placements from the solution space.
However, it is possible that we also removed some good placements or even the best one.
In our opinion, this is not problematic as finding the single best solutions in a solution space of that scale is highly unlikely, and our approach preserves enough good solutions to achieve good results.

\ps{Compare different algorithms}

\begin{figure}[h]
\centering
\captionsetup{justification=centering}
\vspace{-0.5em}
\includegraphics[width=1.0\columnwidth]{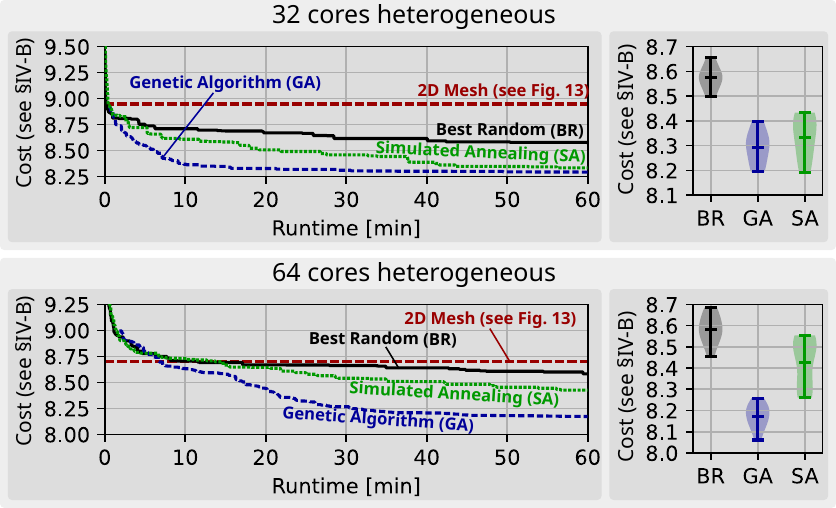}
\caption{\textbf{(\textsection \ref{ssec:hetero-opt}) Results for heterogeneous chiplet shapes.}
We show the evolution of the result over time (left) and the distribution of the final result over 10 repetitions (right).
See \Cref{fig:eval-placements} for the placements found by the best algorithm.}
\label{fig:hetero-results}
\vspace{-1em}
\end{figure}

\Cref{fig:hetero-results} shows our results for the two heterogeneous architectures.
All optimization algorithms outperform the 2D mesh baseline (see \Cref{fig:eval-placements}) and both the \gls{ga} and \gls{sa} outperform \gls{br}.
As in the homogeneous setting, the \gls{ga} performs better than \gls{sa}, however, for the 32 core architecture with heterogeneous chiplet shapes, \gls{sa} performs comparably to the \gls{ga}.

\ps{compare this setting to homogeneous one}

Comparing our results for the heterogeneous setting to those of the homogeneous ones reveals that even though the solution spaces of both settings have the same size, reaching convergence in the heterogeneous setting takes significantly longer than in the homogeneous one.
The reason for this is that the heterogeneous setting is more complex: For each new (order, rotations)-pair that is generated, we need to run our placement algorithm and the function to infer the placement-based \gls{ici} topology.
Therefore, in the heterogeneous setting, the number of placements that an optimization algorithm can create within the time budget is almost an order of magnitude lower than in the homogeneous setting (see \Cref{tab:hetero-ninstance}).

\setcounter{table}{4}
\begin{table}[H]
\setlength{\tabcolsep}{5pt}
\centering
\captionsetup{justification=centering}
\begin{tabular}{lcccc}
\hline
Algorithm					& \makecell{32 cores\\Homog.}		& \makecell{64 cores\\Homog.} 	& \makecell{32 cores\\Heterog.} 	& \makecell{64 cores\\Heterog.}
\vspace{-0.3em}\\
\midrule
Best Random (BR)			& 87.0k 							& 17.3k							& 8.5k							& 1.2k					\\
Genetic Alforithm (GA)		& 41.3k								& 11.5k							& 8.3k							& 1.7k					\\
Simulated Annealing (SA)	& 92.6k								& 20.4k							& 14.1k							& 3.1k					\\
\hline
\end{tabular}
\caption{\textbf{(\textsection \ref{ssec:hetero-opt}) Number of placements generated} by our optimization algorithms for different architectures.}
\label{tab:hetero-ninstance}
\end{table}
\vspace{-1em}
\setcounter{table}{6}

\section{Evaluation}
\label{sec:eval}
\vspace{-0.5em}

\ps{Overview of evaluation section}

\begin{figure*}[t]
\centering
\captionsetup{justification=centering}
\includegraphics[width=1.0\textwidth]{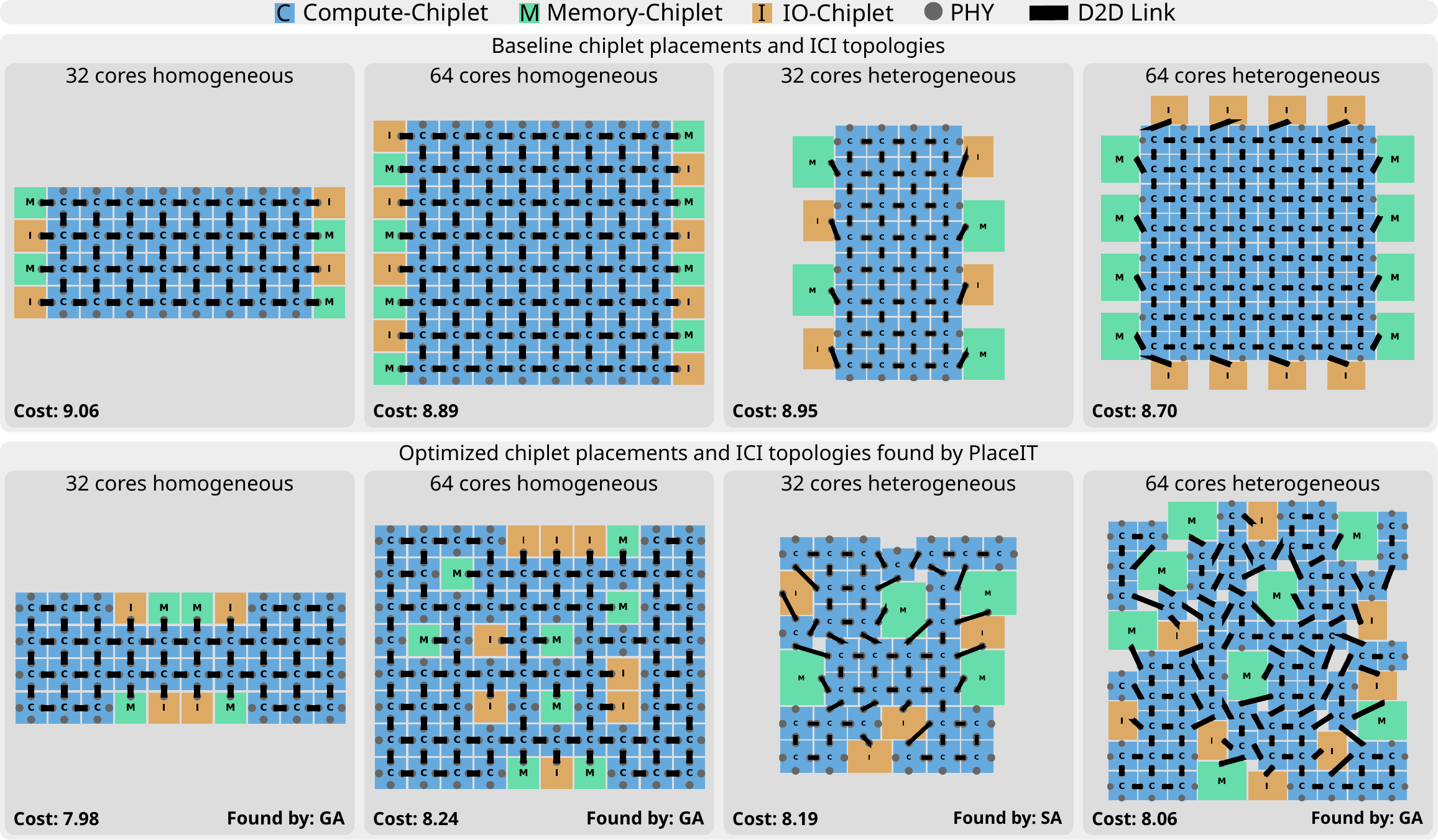}
\caption{\textbf{(\textsection \ref{sec:eval}) Baseline architecture and optimized architecture found by \name~(for the \textit{baseline} configuration)}.}
\label{fig:eval-placements}
\vspace{-1em}
\end{figure*}

We evaluate our proposed chiplet placement and \gls{ici} topology co-optimization methodology on the two homogeneous architectures from \Cref{ssec:homo-opt} and on the two heterogeneous architectures from \Cref{ssec:hetero-opt}.
For each of these four architectures, we design a baseline architecture consisting of a 2D mesh of compute-chiplets in the center with memory- and IO-chiplets on the perimeter.
This type of architecture is the de-facto standard that is used in numerous systems \cite{dataflow_accel_dnn, cifher, simba, hecaton, dojo}.
We perform our evaluation using two different chiplet configurations: 
In the \textit{baseline} configuration, memory- and IO-chiplets only have a single PHY and they cannot relay messages, which is highly unfavorable for \name, as \name~often places memory- and IO-chiplets in the center of the chip (off-chip links of IO-chiplets are routed to the border on the redistribution layer as in AMD's EPYC and Ryzen \cite{amd-chiplet}).
In the \textit{\name} configuration, all chiplets have four \gls{phys} and relay capability.
To ensure a fair comparison, the total memory- and IO-bandwidth stays unchanged and the increased off-chiplet bandwidth due to additional \gls{phys} is only used to relay messages.
\Cref{fig:eval-placements} shows baselines and optimized architectures for the \textit{baseline} configuration.
Unfortunately, a direct comparison to prior work (see \Cref{sec:related-work}) is infeasible since frameworks to optimize the placement are not open-source, or they do not scale to our chiplet counts, and proposals for \gls{ici} topologies on active interposers are not applicable to passive interposers, silicon bridges, and organic substrates.

\vspace{-0.5em}
\subsection{Evaluation Methodology}
\label{ssec:eval-methodology}

\ps{Explain our evaluation methodology and introduce partial trace simulations}

We use RapidChiplet's \cite{rapidchiplet} feature to run simulations in BookSim2 \cite{booksim} using synthetic traffic and application traces from Netrace \cite{netrace}.
BookSim2 is an established, cycle-accurate \gls{noc} simulator and Netrace is a tool for dependency-driven, trace-based \gls{noc} simulations. 
We use the Netrace trace collection \cite{netrace-traces}, which is based on the PARSEC benchmark suite \cite{parsec}.
Each trace is split into five regions (see \Cref{tab:eval-traces}).
Since these traces span across billions of cycles, simulating them in a cycle accurate simulator is extremely time-consuming. 
The \textit{blackscholes\_64c\_simsmall} trace was the only one to terminated within 24 hours, therefore, for the remaining traces, we only simulate the first 1'000'000 cycles of each region.
All traces contain cache coherency traffic between the L1 cache (mapped to compute-chiplets), the L2 cache (mapped to memory-chiplets), and the main memory (mapped to IO-chiplets).

\begin{figure}[H]
\centering
\vspace{-0.5em}
\captionsetup{justification=centering}
\begin{subfigure}{0.99 \columnwidth}
\centering
\includegraphics[width=1.0\columnwidth]{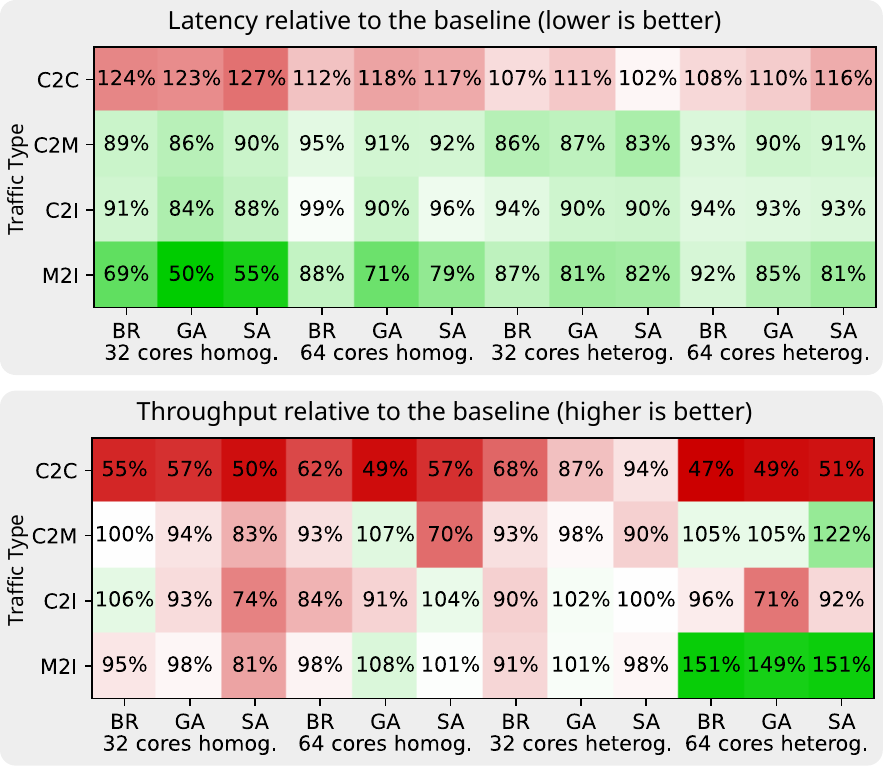}
\end{subfigure}
\caption{\textbf{(\textsection \ref{ssec:eval-synthetic}) Results on synthetic traffic using the \textit{baseline} configuration}.}
\label{fig:eval-synthetic}
\vspace{-1em}
\end{figure}

\begin{figure}[h]
\centering
\vspace{-0.5em}
\captionsetup{justification=centering}
\begin{subfigure}{0.99 \columnwidth}
\centering
\includegraphics[width=1.0\columnwidth]{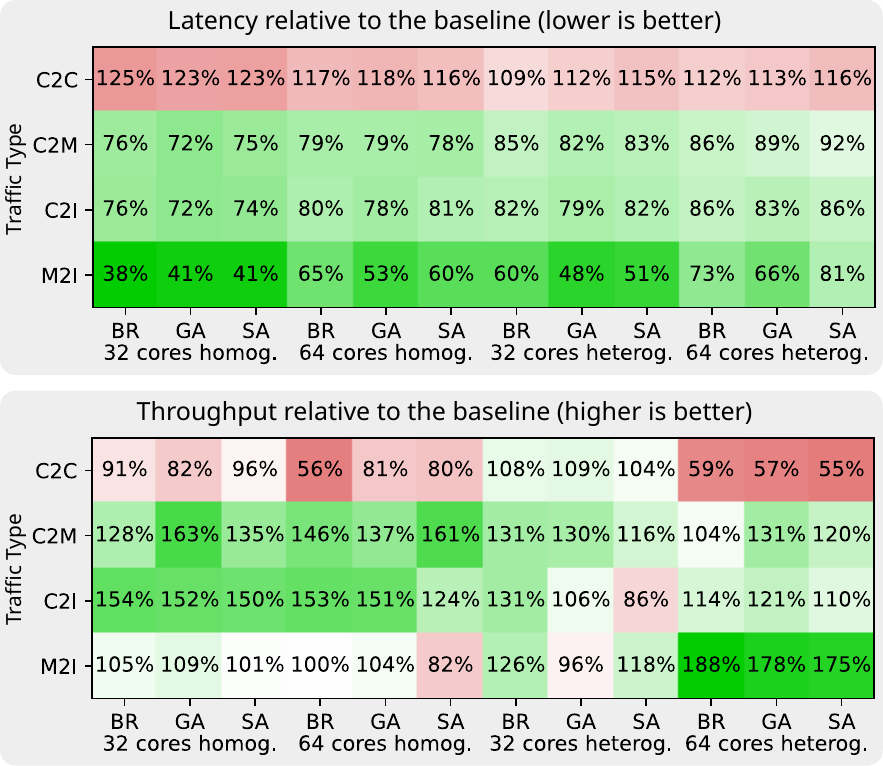}
\end{subfigure}
\caption{\textbf{(\textsection \ref{ssec:eval-synthetic}) Results on synthetic traffic using the \textit{\name} configuration}.}

\label{fig:app-eval-synthetic}
\vspace{-2.2em}
\end{figure}

\ps{Provide remaining simulation details}

We set the parameters of RapidChiplet and BookSim2 to match the latencies described in \Cref{tab:homo-params,tab:hetero-params}. 
BookSim2 models input-queued \gls{vc} routers with a four-stage pipeline (routing, \gls{vc} allocation, switch allocation, crossbar traversal) and wormhole flow control.
We use 1-flit packets for control messages and 9-flit packets for data transfers \cite{netrace-tr}.
Furthermore, we use shortest path routing, up to 8 virtual channels, and 8-flit buffers.

\begin{figure*}[h]
\centering
\captionsetup{justification=centering}
\includegraphics[width=1.0\textwidth]{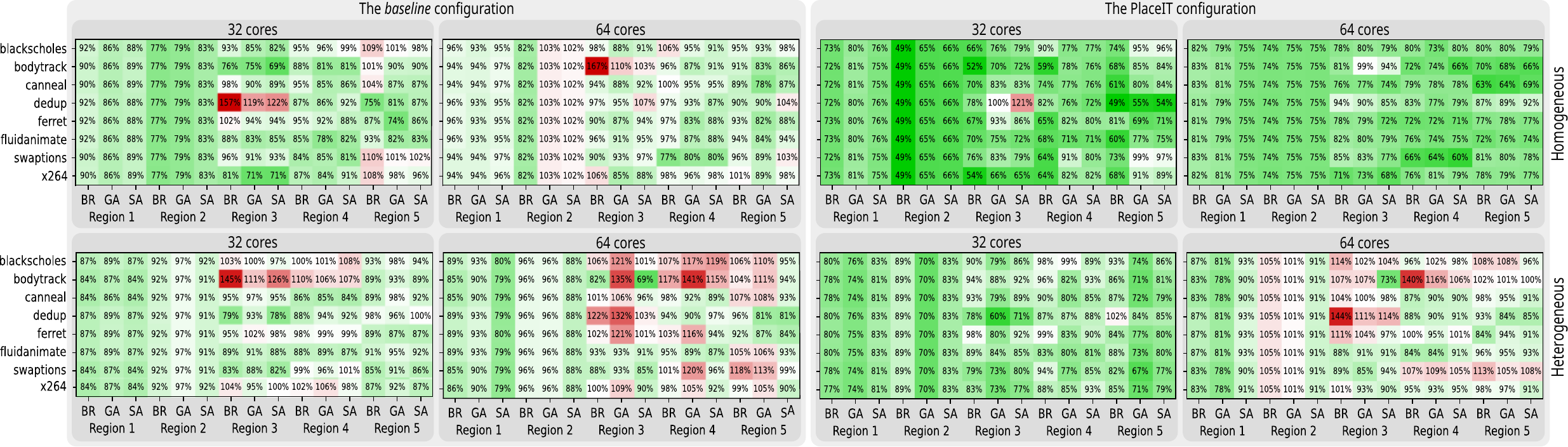}
\caption{\textbf{(\textsection \ref{ssec:eval-trace-partial}) Results for the partial trace regions:} speedup in average packet latency compared to the baseline.}
\label{fig:eval-trace-partial}
\vspace{-1.75em}
\end{figure*}

\subsection{Performance Comparison using Synthetic Traffic}
\label{ssec:eval-synthetic}

\ps{Explain which synthetic traffic we use and why we care about synthetic traffic.}

We compare our optimized \gls{ici} topologies against the baselines in terms of latency and throughput using synthetic \gls{c2c}, \gls{c2m}, \gls{c2i}, and \gls{m2i} traffic.
The advantage of synthetic traffic over real traces is its generality, as synthetic traffic does not depend on the application.
\Cref{fig:eval-synthetic,fig:app-eval-synthetic} show the latency and throughput results under synthetic traffic for the \textit{baseline} and  \textit{\name} chiplet configurations, respectively.

\ps{Discuss results on synthetic traffic: Latency}

Recall that our primary optimization goal was to minimize \gls{c2m} and \gls{m2i} latency and to improve \gls{c2m} and \gls{m2i} throughput.
We observe that for all combinations of architecture and optimization algorithm, \name~improves \gls{c2m}, \gls{c2i}, and \gls{m2i} latency.
The fact that the baseline provides the best \gls{c2c} latency is not surprising, given that in the baseline, compute-chiplets form a regular grid with a 2D mesh topology.

\ps{Discuss results on synthetic traffic: Throughput}

\name~is only able to significantly outperform the baseline architecture in terms of \gls{c2m} and \gls{m2i} throughput if we use the \textit{\name} chiplet configuration, where memory- and IO-chiplets have four \gls{phys} and relay-capabilities. The \textit{baseline} configuration with only a single PHY per memory- and IO-chiplet turns out to be too restrictive to provide significant throughput improvements.

\begin{figure}[h]
\centering
\captionsetup{justification=centering}
\includegraphics[width=1.0\columnwidth]{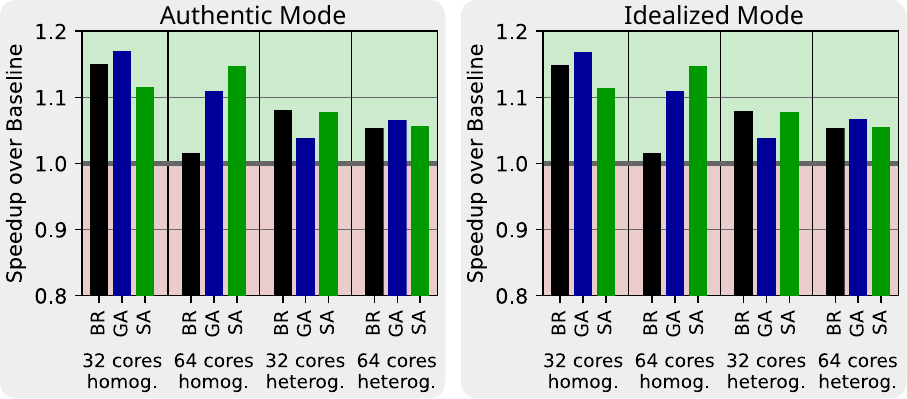}
\caption{\textbf{(\textsection \ref{ssec:eval-trace-full}) speedup over baseline in average packet latency} (blackscholes trace, \textit{baseline} configuration).}
\label{fig:eval-trace-full}
\end{figure}

\begin{figure}[h]
\centering
\captionsetup{justification=centering}
\includegraphics[width=1.0\columnwidth]{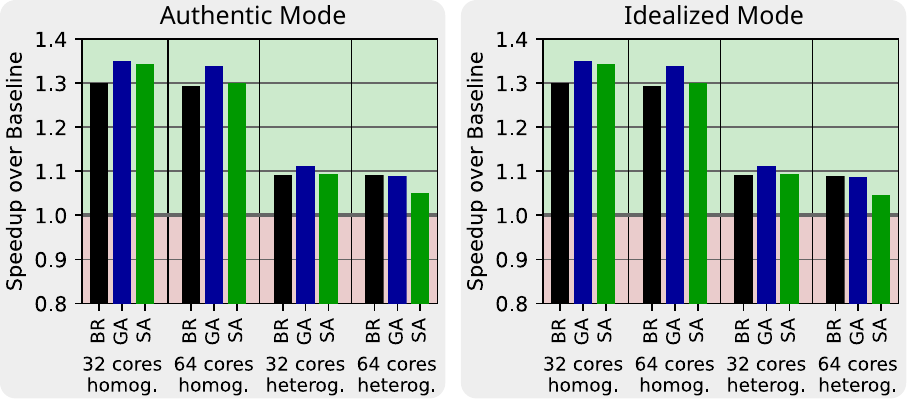}
\caption{\textbf{(\textsection \ref{ssec:eval-trace-full}) speedup over baseline in average packet latency} (blackscholes trace, \textit{\name} configuration).}
\label{fig:app-eval-trace-full}
\end{figure}

\subsection{Performance Comparison on Full Traffic Trace}
\label{ssec:eval-trace-full}

\ps{Explain the two trace modes we use}

We evaluate the performance of our optimized \gls{ici} topologies using the full blackscholes-trace (see \Cref{tab:eval-traces}).
We simulate this trace in two different simulation modes:
In the \emph{authentic} mode, a packet is only injected if all dependencies are satisfied and the cycle, in which the packet appears in the trace, is reached.
This represents a scenario where after receiving a packet, the compute cores need some time to perform computations before injecting the next packet.
The second mode is called \emph{idealized} and it injects a packet as soon as all dependencies are satisfied, assuming ideal cores that perform computations instantly.
This mode is intended as a stress-test for the \gls{ici} as the packet injection rate is significantly higher than in the \emph{authentic} mode.
Our results in \Cref{fig:eval-trace-full,fig:app-eval-trace-full} show that \name~is able to achieve speedups in average packet latency of up to $1.17\times$ (for the \textit{baseline} configuration) and $1.34\times$ (for the \textit{\name} configuration).

\subsection{Performance Comparison on Partial Traffic Traces}
\label{ssec:eval-trace-partial}

\ps{Discuss results on partial traffic traces}

\Cref{fig:eval-trace-partial} shows our results for the simulation of partial trace regions.
\name~is able to reduce the average packet latency to $92\%$ (\textit{baseline} configuration) and $82\%$ (\textit{\name} configuration) on average.
In \Cref{ssec:homo-opt,ssec:hetero-opt} we observed that the \gls{ga} performed significantly better than \gls{br} with respect to the minimization of the cost function.
However, in our partial trace simulation, we see that this is not always the case and in some instances, \gls{br} is even better than the \gls{ga}.
This shows that either our performance estimate or our cost function does not fully reflect the performance on real traces. 
Nevertheless, co-optimizing the chiplet placement and \gls{ici} topology works, as we outperform the baseline architecture in almost all cases.

\subsection{Area Comparison}
\label{ssec:eval-area}

\ps{Compare Area: No area loss compared to manually placed chiplets}

The area of all homogeneous placements for a given architecture is identical, therefore, we only discuss the area of heterogeneous placements. 
For the 32-core architecture, \gls{br} and \gls{sa} increase the area by $5.4\%$ and $0.8\%$ respectively, but the \gls{ga} reduced the area by $8.1\%$ compared to the baseline.
For the 64-core architecture, \gls{br} and \gls{sa} both increase the area by $3.3\%$ but the \gls{ga} reduced the area by $6.3\%$ compared to the baseline.
We conclude that \name~is able to improve the \gls{ici}-performance without introducing significant area overheads.

\begin{table*}[h]
\setlength{\tabcolsep}{6.5pt}
\centering
\captionsetup{justification=centering}
\begin{tabular}{llccclll}
\hline
Name & 
\makecell[l]{Target\\Technology} &
\makecell{Heterogeneous\\Chiplet Shapes} &
\makecell{Optimized\\Placement} & 
\makecell{Optimized\\Topology} & 
\makecell[l]{Target\\Metric} &
\makecell[l]{Optimization\\Method} &
\makecell[l]{Approximate\\Runtime}
\\
\midrule

Ho et al. 			\cite{ho}				& PSI			& \cmark	& \cmark	& \xmark		& TWL, A 		& HB*-tree, SA	& 8 min (4 chiplets)	\\		
Liu et al. 			\cite{liu}				& PSI			& \cmark	& \cmark	& \xmark		& TWL 			& Enumeration, Flow	& 3 h (8 chiplets)	\\		
Seemuth et al. 		\cite{seemuth}			& PSI			& \cmark	& \cmark	& \xmark		& TWL, A 		& SA			& \qmark 				\\		
Eris et al. 		\cite{eris}				& PSI			& \xmark	& \cmark	& \xmark		& P, C 			& Greedy 		& 450 h (16 chiplets) 	\\		
Osmolovskyi et al. 	\cite{osmolovskyi}		& PSI			& \cmark	& \cmark	& \xmark		& TWL 			& B\&B, CSP 	& 1 h  	(11 chiplets) 	\\		
Coskun et al. (1) 	\cite{coskun-1}			& PSI			& \xmark	& \cmark	& (\cmark)$^*$	& P, C, TWL 	& MILP, SA 		& \qmark				\\		
Coskun et al. (2) 	\cite{coskun-2}			& PSI			& \xmark	& \cmark	& (\cmark)$^*$	& P, C, T 		& MILP, SA 		& \qmark 				\\		
Tap-2.5D 			\cite{tap25d}			& PSI			& \cmark	& \cmark	& \xmark		& TWL, T 		& MILP, SA 		& 25 h 	(8 chiplets)  	\\		
Chiou et al. 		\cite{chiou}			& PSI			& \cmark	& \cmark	& \xmark		& TWL, T 		& B\&B, Pruning	& 6 min (11 chiplets)	\\ 		

ButterDonut			\cite{butterdonut}		& ASI			& \xmark	& \xmark	& \cmark		& P 			& Construction	& n/a					\\
ClusCross			\cite{cluscross}		& ASI			& \xmark	& \xmark	& \cmark		& P 			& Construction	& n/a					\\
Kite 				\cite{kite}				& ASI			& \xmark	& \xmark	& \cmark		& P 			& Construction	& n/a					\\

HexaMesh			\cite{hexamesh}			& OPS, PSI		& \xmark	& \cmark	& \cmark		& P 			& Construction	& n/a					\\
\midrule
\name~(This Work)							& OPS, SB, PSI	& \cmark	& \cmark	& \cmark		& P, A, (TWL)	& SA, GA 		& 1 h (73 chiplets)	\\
\hline
\end{tabular}
\caption{\textbf{(\textsection \ref{sec:related-work}) Overview of related work.}
\underline{Target technology}: \textbf{OPS}: Organic package substrate, \textbf{SB}: Silicon bridge, \textbf{PSI}: Passive silicon interposer, \textbf{ASI}: Active silicon interposer.
\underline{Target metric}: \textbf{TWL}: Total wire length, \textbf{T}: Temperature, \\\textbf{P}: Performance, \textbf{C}: Cost, \textbf{A}: Area.
\underline{Optimization method}: \textbf{SA}: Simulated annealing, \textbf{B\&B}: Branch \& bound, \textbf{CSP}: Constraint-\\satisfaction problem, \textbf{MILP}: Mixed integer-linear problem, \textbf{GA}: Genetic algorithm. 
\textbf{$^*$}Select best topology out of 8 candidates.
}
\label{tab:related-work}
\end{table*}
\section{Related Work}
\label{sec:related-work}

\ps{Optimize placement for existing topology}

Multiple recent studies have focused on placing chiplets on a passive silicon interposer.
These works usually assume that the \gls{ici} topology is given as an input.
Most of them optimize the \gls{twl} \cite{ho, liu, seemuth, osmolovskyi} or a combination of \gls{twl} and thermal properties \cite{chiou, tap25d}.
Some also consider the \gls{ici} performance or the cost \cite{eris}.
Many of these works can be combined with \name. 
We could, e.g., use \name~to find a placement and \gls{ici} topology, and then apply TAP-2.5D \cite{tap25d} to fine-tune the placement for thermal properties. 

\ps{Explain Coskun in detail, since they kind of optimize the topology}

An interesting line of work is that of Coskun et al. \cite{coskun-1, coskun-2}.
They apply a cross-layer co-optimization approach to jointly optimize a 2.5D stacked chip across the logical-, physical- and circuit layer.
They consider a predetermined set of well-known \gls{ici} topologies out of which they select the most suitable one.
This is in contrast to \name, where completely new \gls{ici} topologies are created.
We see potential in combining the two approaches by first finding an \gls{ici} topology and placement using \name~and then applying the cross-layer co-optimization approach to optimize the remaining layers or to select the placement found by either \gls{br}, the \gls{ga} or \gls{sa}.

\ps{Optimized topologies for active interposers}

Research on \gls{ici} topologies focuses on active interposers, since they offer longer links and package-level routers.
Such works assume a regular 2D grid of compute-chiplets with memory- or IO-chiplets on the side.
\gls{ici} topologies such as ButterDonut \cite{butterdonut}, ClusCross \cite{cluscross}, or Kite \cite{kite} are optimized for low \gls{ici} latency and high \gls{ici} throughput.

\ps{HexaMesh and its shortcomings}

One of the few works focussing on \gls{ici} topologies for passive silicon interposers is HexaMesh \cite{hexamesh}.
They propose a hexagonal arrangement of chiplets where each non-border chiplet has six \gls{d2d} links to other chiplets.
However, this approach is only applicable to homogeneous architectures.

\ps{Explain how our approach fills a gap}

\name~is the first work known to us that jointly optimizes \gls{ici} topology and chiplet placement.
Furthermore, it is the first work on \gls{ici} topologies for heterogeneously shaped chiplets.
Table \ref{tab:related-work} compares \name~to its related work.

\section{Conclusion}
\label{sec:conclusion}
\vspace{0.7em}

\ps{Highlight key idea/novelty of PlaceIT}

In this work, we present \name, a novel methodology to jointly optimize the chiplet placement and \gls{ici} topology for chips with heterogeneous chiplet shapes and silicon bridges or passive silicon interposers.
The main novelty of our approach is that we perform optimization on the chiplet placement itself, where we infer a custom, placement-based \gls{ici} topology for each placement produced by an optimization algorithm.
We use the placement and its inferred \gls{ici} topology to compute proxies for \gls{ici} latency and throughput of different traffic types, which we combine into a user-defined quality metric that is returned to the optimization algorithm.

\ps{Highlight the framework/code}

The open-source \name~framework is modular and allows adding custom optimization algorithms or placement representations.
\name~offers a wide range of configurable parameters, making it applicable for a variety of designs with different chiplet dimensions, PHY-counts, and \gls{d2d} links.

\ps{Summarize evaluation}

Our evaluation on synthetic traffic shows that \name~produces \gls{ici}s with vastly lower \gls{c2m}, \gls{c2i}, and \gls{m2i} latency (reduced by up to 62\%) compared to a 2D mesh baseline.
On real traffic traces, \name~reduces the average packet latency in almost all traces and architectures considered.
The average packet latency is reduced by up to $18\%$ on average.

\ps{Concluding sentence}

By using our open-source \name~framework, architects can co-optimize their chiplet-placement and \gls{ici} topology to build 2.5D stacked chips with low-latency interconnects.

\ifnb
\section*{Acknowledgements}
\label{sec:ack}

\ps{ACKs - Timo's OK is pending}

This work was supported by the ETH Future Computing Lab (EFCL), financed by a donation from Huawei Technologies.
It also received funding from the European Research Council
\raisebox{-0.25em}{\includegraphics[height=1em]{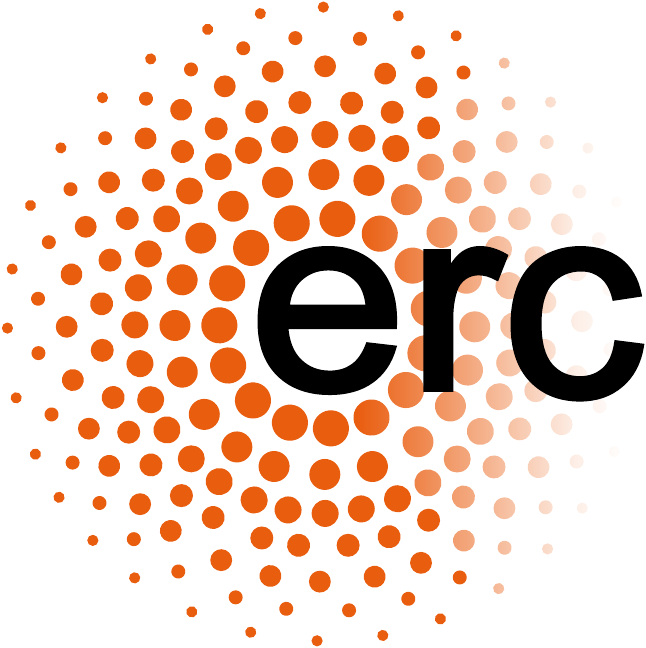}} (Project PSAP,
No.~101002047) and from the European Union's HE research 
and innovation programme under the grant agreement No.~101070141 (Project GLACIATION).

\fi

\newpage
\appendices
\crefalias{section}{appendix}
\crefalias{subsection}{appendix}


\newpage
\bibliographystyle{IEEEtranS}
\bibliography{bibliography}

\end{document}